\documentclass[twocolumn]{aastex631}
\usepackage{amsmath}
\usepackage{hyperref}
\usepackage{multirow}
\usepackage{makecell}
\usepackage{rotating}
\usepackage{booktabs}
\usepackage{subfigure}
\usepackage{graphicx}

\begin{document}


\title{On the Difference Between Pulsar Radio Emission Beams from the Two Poles}

\defcitealias{Wang+2014apj}{WPZ14}
\defcitealias{Johnston+2019bmnras}{JK19}
\defcitealias{Blaskiewicz+1991apj}{BCW}
\defcitealias{Radhakrishnan+1969apl}{RC69}

\author[0000-0002-9119-1388]{Xiancong Wu}
\affiliation{Department of Astronomy, School of Physics and Materials Science, Guangzhou University, Guangzhou 510006, People's Republic of China}
\affiliation{National Astronomical Data Centre, the Great Bay Area, Higher Education Mega Centre, Guangzhou 510006, People's Republic of China}
\email{sanderng@e.gzhu.edu.cn}

\author[0000-0002-2044-5184]{Hongguang Wang}
\affiliation{Department of Astronomy, School of Physics and Materials Science, Guangzhou University, Guangzhou 510006, People's Republic of China}
\affiliation{National Astronomical Data Centre, the Great Bay Area, Higher Education Mega Centre, Guangzhou 510006, People's Republic of China}
\email{hgwang@gzhu.edu.cn}

\author[0000-0001-7120-4076]{Hao Tong}
\affiliation{Department of Astronomy, School of Physics and Materials Science, Guangzhou University, Guangzhou 510006, People's Republic of China}
\affiliation{National Astronomical Data Centre, the Great Bay Area, Higher Education Mega Centre, Guangzhou 510006, People's Republic of China}

\author[0000-0002-4300-121X]{Rui Luo}
\affiliation{Department of Astronomy, School of Physics and Materials Science, Guangzhou University, Guangzhou 510006, People's Republic of China}
\affiliation{National Astronomical Data Centre, the Great Bay Area, Higher Education Mega Centre, Guangzhou 510006, People's Republic of China}

\author[0000-0002-6437-0487]{PengFei Wang}
\affiliation{National Astronomical Observatories, Chinese Academy of Sciences, Beijing 100101, People's Republic of China}
\affiliation{School of Astronomy, University of Chinese Academy of Sciences, Beijing 100049, People's Republic of China}

\author[0009-0003-8221-9611]{Chengbing Lyu}
\affiliation{Department of Astronomy, School of Physics and Materials Science, Guangzhou University, Guangzhou 510006, People's Republic of China}
\affiliation{National Astronomical Data Centre, the Great Bay Area, Higher Education Mega Centre, Guangzhou 510006, People's Republic of China}

\author[0009-0005-1848-0553]{Hai Lei}
\affiliation{Department of Astronomy, School of Physics and Materials Science, Guangzhou University, Guangzhou 510006, People's Republic of China}
\affiliation{National Astronomical Data Centre, the Great Bay Area, Higher Education Mega Centre, Guangzhou 510006, People's Republic of China}








\begin{abstract}

The long-standing assumption of symmetric radio emission beams from the two magnetic poles of pulsars is challenged by observational evidence of asymmetry and underfill. Direct testing of this symmetry remains difficult for most pulsars. As an indirect test, we collected polarization profiles of 11 interpulse pulsars observed with the Five-hundred-meter Aperture Spherical radio Telescope, MeerKAT, and Parkes. We developed a rotating vector model incorporating aberration and retardation effects to fit the position angle swings of selected pulsars, thereby determining the intrinsic emission region corresponding to the observed pulse windows. Based on both the conal and fan beam models, we compared three key parameters—beam radius, magnetic azimuth width, and emission intensity—between the intrinsic emission regions of the main pulse and interpulse. Among the eight pulsars with a confirmed double-pole geometry, none exhibits similarity in the azimuth width. Only two show potentially similar beam radii, while six demonstrate comparable emission intensities within specific parameter spaces. These results indicate that the emission beams from the two magnetic poles of a pulsar may be generally dissimilar in size, suggesting that the physical conditions governing pair production and particle acceleration differ between the two poles. The random distribution of active emission regions further implies inhomogeneity within the polar cap, which may originate from the differences in local magnetic field structure or surface properties.

\end{abstract}



\section{Introduction} \label{sec:intro}

Pulsars are commonly assumed to possess a large-scale dipole magnetic field, with identical gap structures at both magnetic poles where charged particles are accelerated and emit radio emission \citep{Goldreich+1969apj, Ruderman+1975apj, Dyks+2003apj, Qiao+2004apjl}. This symmetry implies that the distributions of secondary pairs across the polar cap should also be identical at both poles, resulting in identical radio emission beams. Although widely adopted, this simple and attractive model may not accurately reflect the real conditions in pulsars. Several observations have revealed either underfilling of emission beams or asymmetries in the double-pole emission from a few pulsars, raising the question of whether dissimilar radio emission beams at the two magnetic poles are a common feature.

The most compelling evidence for such an asymmetry comes from PSR J1906$+$0746, a precessing pulsar in a binary system \citep{Lorimer+2006apj}. With a nearly orthogonal emission geometry, this pulsar allows observations of radio emission from both magnetic poles. Its precession causes the line of sight (LOS) to sweep across different parts of the radio emission beams over time. By modeling the multiepoch polarization position angle (PA) swings using the rotating vector model \citep[RVM; ][hereafter \citetalias{Radhakrishnan+1969apl}]{Radhakrishnan+1969apl}, \cite{Desvignes+2019sci} reconstructed the radio emission beam patterns for both poles and found significant differences in the beam shape, intensity profile, and polarization profile. Even in the regions with potential communication between the two poles, the beamwidth, intensity profile, and polarization profile differ considerably.

However, this pulsar remains the only direct test case among thousands of known pulsars. Other binary pulsars with significant precessional effects—e.g., PSRs B1913$+$16 \citep{Hulse+1975ApJ, Weisberg+2002apj}, B1534$+$12 \citep{Wolszczan1991Natur, Arzoumanian+1996apj}, J1141$-$6545 \citep{Kaspi+2000ApJ, Manchester+2010apj}, J0737$-$3039A/B \citep{Lyne+2004Sci, Perera+2012apj, Perera+2014apj, Lomiashvili+2014mnras, Saha+2017MNRAS}, and J1946$+$2052 \citep{Stovall+2018ApJ, Meng+2024apj}—either exhibit PA swings that deviate from the RVM and/or lack interpulse emission, making it difficult to reliably determine the emission geometry and hence reconstruct their emission beams. For isolated pulsars without precession, the fixed viewing angle allows only a constant slice of the emission beam to be observed, preventing the reconstruction of the emission beams and direct comparison of the beams from the two poles.

Although direct testing is challenging, an indirect test is possible for pulsars that show two-pole-origin interpulse emission together with an S-shaped PA swing. Once the inclination angle $\alpha$ and impact angle $\beta$ are inferred from the RVM fit to the PA swing, the emission loci for both magnetic poles can be determined. Assuming the two poles share the same beam morphology, the beam parameters can be constrained for each pole. These parameters can then be compared to assess the similarity of the emission beams. Previous studies on interpulse pulsars have focused on population properties and beam evolution \citep{Weltevrede+2008mnras, Maciesiak+2011mnras, Arzamasskiy+2017mnras}, as well as on the beam structure and emission regions \citep[e.g.,][hereafter \citetalias{Johnston+2019bmnras}]{Kramer+2008mnras, Weltevrede+2009mnras, Keith+2010mnras, Sun+2025apj, Johnston+2019bmnras}. Asymmetry in emission regions and the underfilling of the polar cap have attracted attention in recent studies. For instance, in a study of five interpulse pulsars, \cite{Keith+2010mnras} found that, for two of them, either the emission originates from ``closed'' field lines or the emission regions are highly asymmetric with respect to the meridian plane\footnote{The plane containing the rotation and the magnetic axes}. However, no systematic study has yet investigated the similarity of the emission beams at the two magnetic poles across a sample of interpulse pulsars.

Such a study requires emission beam models that provide quantitative relationships between beam morphology parameters and observable pulse profile parameters, such as the pulse width and the emission intensity. Currently, three types of emission beam models exist: the conal beam model \citep{Ruderman+1975apj, Rankin1983apj}, the patchy beam model \citep{Lyne+1988mnras, 1995JApA...16..107M}, and the fan beam model \citep[][hereafter \citetalias{Wang+2014apj}]{Dyks+2010mnras, Wang+2014apj}. In the conal beam model, the beam shape is usually assumed circular, with beam radius related to the observed pulse width, $\alpha$, and $\beta$. In the patchy beam model, the active emission regions randomly distributed within the polar cap. Lacking distinct boundaries, this model cannot provide quantitative morphology parameters. In the \citetalias{Wang+2014apj} fan beam model, the azimuthal width of an active magnetic flux tube is a key parameter that determines the angular width of the emission beam and can also be related to $\alpha$, $\beta$, and the pulse width. The \citetalias{Wang+2014apj} model also provides a quantitative radial intensity profile, allowing us to derive the emission beam intensity from the observed pulse intensity and the corresponding emission radius. Therefore, given an emission beam model, the beam parameters can be constrained from the observed pulse width, $\alpha$, $\beta$, and flux density, enabling a comparison of the emission beams from the two magnetic poles.

This paper is organized as follows. After selecting pulse profiles from diverse datasets based on specific criteria (Section \ref{sec:data}), we introduce an aberration and retardation (A/R)-corrected RVM to fit the observed PA swings (Section \ref{subsec:ar}). The intrinsic emission longitudes corresponding to each observed pulse longitude are then obtained from the fitted results and the observed pulse windows (Section \ref{subsec:geo} and \ref{subsec:emregion}). With the intrinsic emission longitudes determined for both poles of the selected pulsars, we compare their emission beams within the parameter space provided by the conal beam model (Section \ref{subsec:conal}) and the fan beam model (Section \ref{subsec:fanbeam}). The results are summarized and discussed in Section \ref{sec:psrs} and \ref{sec:conc&diss}.

\section{Data acquisition}
\label{sec:data}

\begin{figure*}
    \centering
    \includegraphics[width=\linewidth]{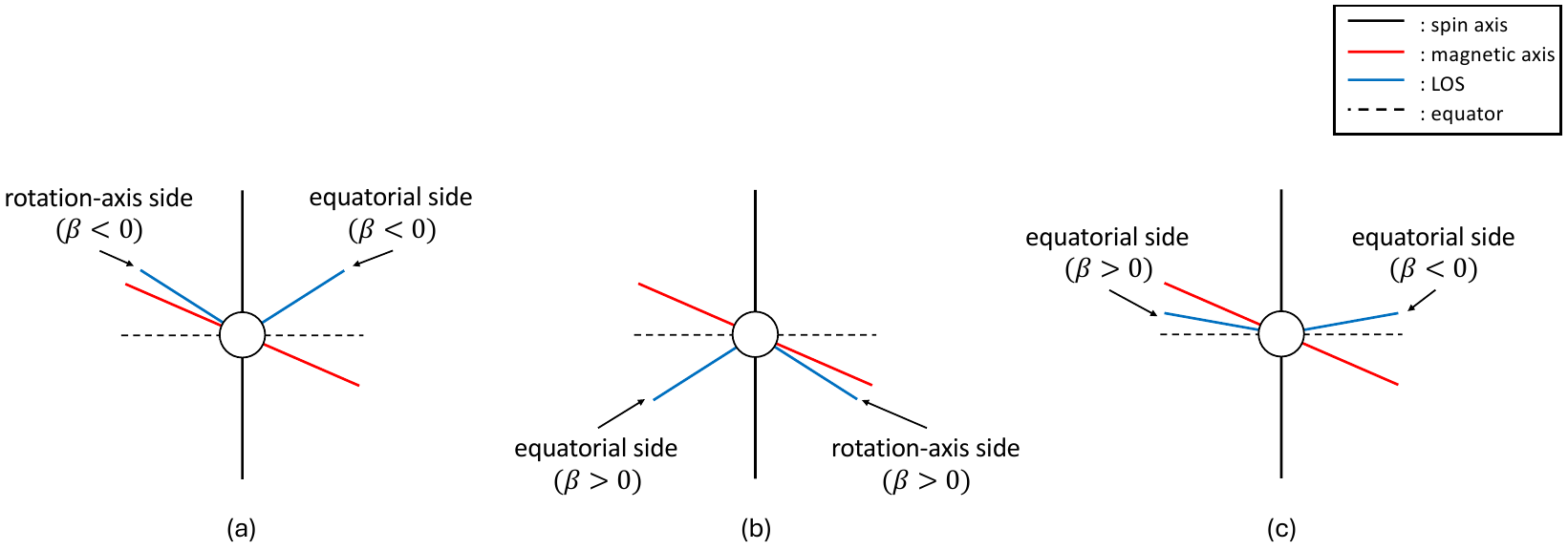}
    \caption{Schematic illustration of whether the main pulse and interpulse originate from different sides or the same side of the two magnetic poles. The neutron star is depicted with black circles. Red, blue, and black solid lines represent the magnetic axis, LOS, and rotational axis, respectively. The black dashed lines denote the neutron star equator. Panels (a) and (b) show cases where the main pulse and interpulse emission arise from different sides of their respective poles, while panel (c) illustrates the case where the main pulse and interpulse emission originate from the same side (the ``equatorial side'') of their poles.}
    \label{fig:ontheside}
\end{figure*}

To investigate the similarity of radio emission beams from the two poles of interpulse pulsars, we apply the following selection criteria:
\begin{enumerate}
    \item [(1)]
    Millisecond pulsars possess very compact magnetospheres, where emission can originate from diverse loci, often leading to complex pulse profiles and PA swings that deviate from the RVM \citep{Kramer+2025arXiv}. We therefore restrict our sample to normal pulsars.
    
    \item [(2)]
    Reliable determination of the emission geometry requires that the PA swings be well fitted by the RVM. We therefore select sources with regular S-shaped PA swings.
    
    \item [(3)]
    The interpulse flux density must exceed a $3\sigma$ detection threshold to ensure accurate measurement of its pulse window and emission intensity.
    
    \item [(4)]
    The main pulse and interpulse should be separated by approximately $180^\circ$ in pulse longitude. In addition, steep PA swings at both the main pulse and the interpulse are required to confirm a nearly orthogonal rotator, for which our LOS cuts across both magnetic poles.
    
    \item [(5)]
    Another factor to consider is the possible asymmetry in the distribution of net charge number density above the polar cap for inclined rotators. This arises primarily from the difference in $\mathbf{\Omega}\cdot\mathbf{B}$ between the region on the side of the magnetic pole toward the spin axis (hereafter the ``rotation-axis side'') and the region on the side toward the equator (hereafter the ``equatorial side''), where $\mathbf{\Omega}$ is the angular velocity, and $\mathbf{B}$ is the magnetic field, as seen in some simulations \citep[e.g.,][]{Bai+2010apj, Gralla+2017apj}. Consequently, the emission beams on the rotation-axis side and on the equatorial side may differ. To minimize this effect, we select only interpulse pulsars for which the LOS samples the same side of both poles.
\end{enumerate}

To implement criterion (5), we first define the inclination and impact angles for the interpulse:
\begin{equation} \label{equ:alpha_trans}
    \alpha_{\rm I} = 180^\circ - \alpha_{\rm M}
\end{equation}
and
\begin{equation} \label{equ:beta_trans}
    \beta_{\rm I}=2\alpha_{\rm M}+\beta_{\rm M}-180^\circ \,,
\end{equation}
where the subscripts ``M'' and ``I'' denote the main pulse and interpulse, respectively. 
As shown in Figure \ref{fig:ontheside}, the observed side of the magnetosphere can then be defined consistently for the two poles: 
\begin{itemize}
    \item 
    If $\alpha < 90^\circ$ and $\beta > 0^\circ$, or $\alpha > 90^\circ$ and $\beta < 0^\circ$, the LOS sweeps across the equatorial side.
    \item 
    If $\alpha < 90^\circ$ and $\beta < 0^\circ$, or $\alpha > 90^\circ$ and $\beta > 0^\circ$, the LOS sweeps across the rotation-axis side.
\end{itemize}

If the LOS sweeps across the same side of the magnetosphere at both poles, it must be the equatorial side (panel (c) in Figure \ref{fig:ontheside}). This geometry requires $\beta_{\rm M}$ and $\beta_{\rm I}$ to have opposite signs. Furthermore, because the maximum slope of the PA swing is given by
\begin{equation}
    \left(\frac{\rm d \psi}{\rm d \phi}\right)_{\rm max}=\frac{\sin \alpha}{\sin \beta} \,,
\end{equation}
the slopes of the PA swings for the main pulse and interpulse must also have opposite signs in this case. This opposite-slope criterion serves as a practical indicator for selecting pulsars that satisfy criterion (5).

Based on these criteria, integrated polarization pulse profiles are selected from three major open-source databases.
First, we use data from the Five-hundred-meter Aperture Spherical radio Telescope (FAST). Through the Galactic Plane Pulsar Snapshot (GPPS) survey,\footnote{\url{http://zmtt.bao.ac.cn/GPPS/}} \cite{Han+2021raa, Han+2025raa} have discovered 751 new pulsars to date. High-quality integrated polarization profiles of both newly discovered and previously known pulsars at the L band from this survey and follow-up projects are compiled in the FAST pulsar database\footnote{\url{http://zmtt.bao.ac.cn/psr-fast/}} by \cite{Wang+2023raa}. 
Second, for southern-sky pulsars, we incorporated data from the Thousand Pulsar Array program conducted with MeerKAT. Initiated by \cite{Johnston+2020mnras} in early 2019 to study radio emission from normal pulsars, this project has released 1271 polarization pulse profiles at the L band after about 3 yr of observations \citep{Posselt+2023mnras}. These integrated pulse profiles are available online \citep{https://doi.org/10.5281/zenodo.7272361}.
The third database consists of archival observations from the Parkes telescope dating back to late 2004. \cite{Johnston+2018mnras} have reprocessed these full-polarization data of 600 normal pulsars observed at the L band and published the resulting integrated pulse profiles.
Basic information for the 11 selected pulsars is listed in Table \ref{tab:para}. When multiple telescopes observed the same pulsar, the highest-quality pulse profile was chosen for this study.

\begin{deluxetable}{ccccc}
\tablecaption{Information for the selected pulsars. \label{tab:para}}

\tablehead{
\colhead{Telescopes} & \colhead{PSRJ} & \colhead{Period} & \colhead{Sep.} & \colhead{Ref.} \\
\colhead{} & \colhead{} & \colhead{($\rm ms$)} & \colhead{($^\circ$)} & \colhead{}}

\startdata
\multirow{3}{*}{FAST}
& 1913$+$0832 & 134 & 192 &
\multirow{3}{*}{WHX23} \\
& 2047$+$5029 & 446 & 181 & \\
& 2208$+$4056 & 637 & 182 & \\
\hline
\multirow{5}{*}{MeerKAT}
& 1126$-$6054 & 203 & 186 & 
\multirow{5}{*}{PKJ23} \\
& 1413$-$6307 & 395 & 173 & \\
& 1611$-$5209 & 182 & 176 & \\
& 1755$-$0903 & 191 & 167 & \\
& 1909$+$0749 & 237 & 187 & \\
\hline
\multirow{3}{*}{Parkes}
& 0908$-$4913 & 107 & 173 & 
\multirow{3}{*}{JK18} \\
& 1549$-$4848 & 288 & 176 & \\
& 1739$-$2903 & 323 & 183 & \\
\enddata
\tablecomments{Column (1): telescopes used. Column (2):﻿ pulsar name. Column (3): period. Column (4): separation between the main pulse and the interpulse. Column (5): references—WHX23: \cite{Wang+2023raa}; PKJ23: \cite{Posselt+2023mnras}; JK18: \cite{Johnston+2018mnras}.}
\end{deluxetable}

\section{Methodology for determining the intrinsic emission longitudes} \label{sec:meth}

Radio emission from pulsars is thought to originate from a certain altitude above the stellar center, causing the observed emission longitudes to differ from the intrinsic longitudes. This discrepancy arises primarily from two effects. Aberration shifts the emission in the direction of rotation, introducing a phase advance. Retardation, due to the finite travel time from higher altitudes, also causes a phase advance. Together, A/R effects shift the observed pulse longitudes ahead of their intrinsic values. To compare the emission beams from the two poles of selected pulsars, we must first correct for the A/R effects and obtain the intrinsic emission regions at both poles.

Previous studies have often used the first-order approximation
\begin{equation}
    \label{equ:bcwshift}
    \Delta \phi = \frac{4 h_{\rm em}}{R_{\rm LC}}
\end{equation}
derived by \cite{Blaskiewicz+1991apj} to correct the A/R phase shifts, where $\Delta \phi$, $h_{\rm em}$, and $R_{\rm LC} = cP / 2\pi$ are the A/R phase shift, emission height, and light cylinder radius, respectively. The phase difference between the centroid of the pulse profile and the fiducial phase derived from the PA swing is commonly used with this formula to infer the emission height \citep[e.g.,][\citetalias{Johnston+2019bmnras}]{Rookyard+2015mnras}. However, this approximation assumes a fully filled, symmetric emission region. In asymmetric cases, due to the different emission heights at the leading and trailing boundaries, the measured phase difference may not reflect the true emission height, and the shifted pulse longitudes may not be the intrinsic ones. Moreover, for pulsars with $\Delta \phi < 0$, this method is no longer applicable \citep[e.g.,][]{Johnston+2023MNRAS, Wang+2023raa}. Alternative methods estimate phase shifts separately at the leading and trailing boundaries \citep[e.g.,][]{Gangadhara+2001apj, Weltevrede+2009mnras, Keith+2010mnras}, offering better handling of asymmetry. However, a more comprehensive approach is desirable.

In this work, we introduce a numerical A/R-corrected RVM by applying the A/R effects to the intrinsic RVM (i.e., the figure presented by \citetalias{Radhakrishnan+1969apl}). This involves modeling the magnetic field (Section \ref{subsubsec:field}) and calculating the aberration phase shift using the Lorentz transformation and the retardation phase shift via propagation time (Section \ref{subsubsec:abe&ret}). In this model, a footprint parameter, $\eta$, is introduced to describe different layers of open field lines ($\eta = 1$ corresponds to the last open field line, $\eta = 0$ to the magnetic axis). The A/R-corrected emission longitudes, representing the observed pulse longitudes, are then related to the intrinsic emission longitudes. By fitting the observed PA swing using the A/R-corrected RVM (Section \ref{subsec:geo}), the intrinsic emission longitudes corresponding to each observed pulse longitude can be obtained by subtracting the A/R phase shifts (Section \ref{subsec:emregion}).

\subsection{A/R-corrected RVM}
\label{subsec:ar}

This section details the numerical A/R-corrected RVM. We first model the magnetic field to obtain the position vectors $\mathbf{r}$ and the emission directions $\mathbf{k}$ (tangent to the $\mathbf{B}$ field) for each intrinsic emission longitude. The Lorentz transformation is then applied to the emission directions to obtain the aberrated emission direction. Assuming the electric vectors of the emissions lie in the $\mathbf{k}-\mathbf{B}$ plane, the intrinsic and aberrated PA swings are calculated. Finally, using the position vectors and aberrated emission directions, the retardation phase shift is determined from the propagation time from the stellar center to the emission point. To simplify the model, and because the radio emission is thought to originate at relatively low altitudes, we ignore magnetic sweepback and perform calculations for a static inclined dipole.

\begin{figure}[ht]
    \centering
    \includegraphics[width=\linewidth]{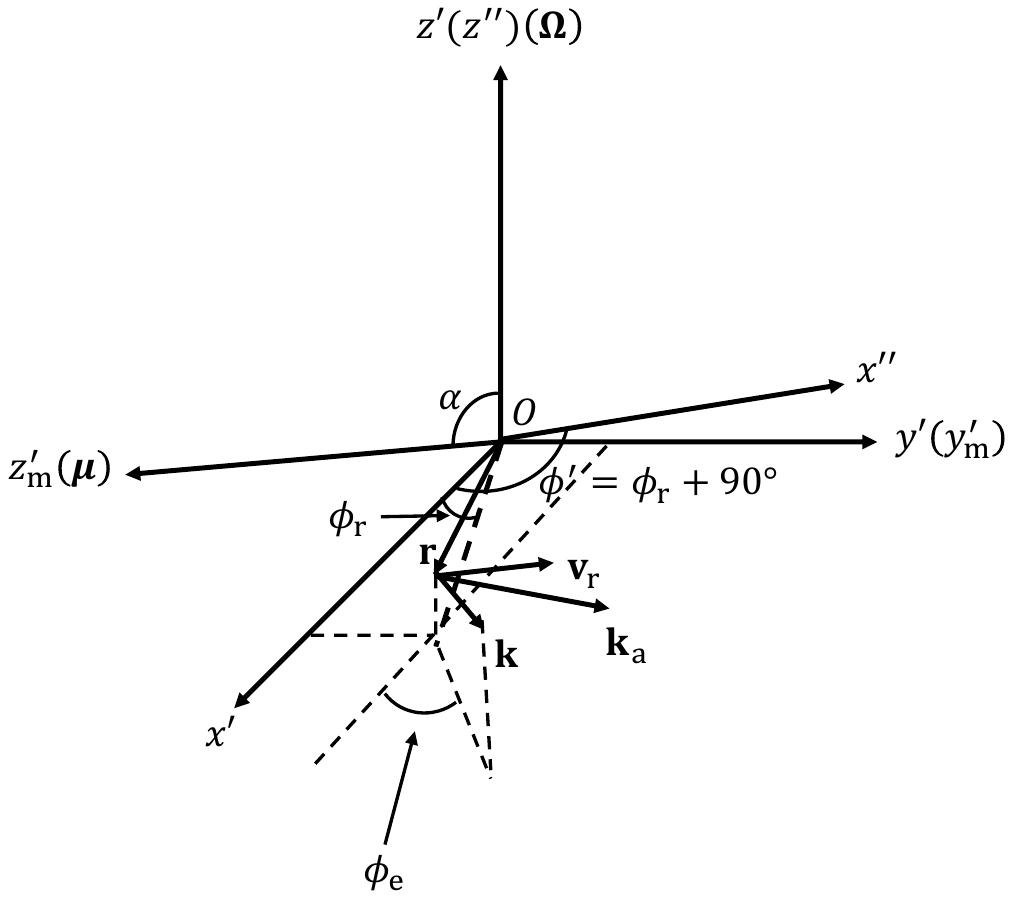}
    \caption{Schematic diagram of three coordinate systems. The $Ox'y'z'$ system is $\mathcal{K}'$, while $Ox_{\rm m}'y_{\rm m}'z_{\rm m}'$ is $\mathcal{K}_{\rm m}$; both are in the corotating frame. The $x'Oz'$ and $x_{\rm m}'Oz_{\rm m}'$ planes lie in the meridian plane, with an angle $\alpha$ between the $z_{\rm m}'$- and $z'$-axes. The $Ox''y''z''$ system is $\mathcal{K}''$, with the $x''$-axis parallel to the instantaneous rotational velocity $\mathbf{v}_{\rm r}$. The longitude of $\mathbf{r}$ is $\phi_{\rm r}$. Vector $\mathbf{k}$ is the emission direction, with longitude $\phi_{\rm e}$. The vectors $\mathbf{v}_{\rm r}$ and $\mathbf{k}_{\rm a}$ are the rotational velocity and aberrated emission direction, respectively.}
    \label{fig:coordsys}
\end{figure}

As illustrated in Figure \ref{fig:coordsys}, we define three coordinate systems:
\begin{itemize}
    \item The magnetic coordinate system $\mathcal{K}'_{\rm m}$ in the corotating frame, where the $z'_{\rm m}$-axis aligns with the magnetic dipole moment $\boldsymbol{\mu}$;
    \item The coordinate system $\mathcal{K}'$ in the corotating frame, with the rotation axis $\boldsymbol{\Omega}$ as its $z'$-axis;
    \item The instantaneous inertial coordinate system $\mathcal{K}''$, where its $z''$-axis aligns with $\boldsymbol{\Omega}$ and its $x''$-axis points in the direction of the instantaneous rotational velocity vector $\mathbf{v}_{\rm r}$.
\end{itemize}
The $x'_{\rm m}Oz'_{\rm m}$ and $x'Oz'$ planes lie in the meridian plane. The angle between the $z'$-axis and the $z'_{\rm m}$-axis is the inclination angle $\alpha$. Henceforth, superscripts/subscripts on vectors indicate the coordinate system in which they are expressed.

\subsubsection{Static dipole}
\label{subsubsec:field}

The following calculations are based on an inclined dipole with an inclination angle $\alpha$ and an LOS impact angle $\beta$. For a given intrinsic emission longitude $\phi_{\rm e}$, the emission radius $\rho'_{\rm m}$ (the angle between $\mathbf{k}'_{\rm m}$ and the magnetic axis) and the magnetic azimuth angle $\varphi'_{\rm m}$ of $\mathbf{k}'_{\rm m}$ can be derived from spherical geometry: 
\begin{equation}
    \label{equ:calrho}
    \cos\rho’_{\rm m}=\cos\alpha \cos(\alpha +\beta) + \sin\alpha \sin(\alpha+\beta) \cos\phi_{\rm e}
    \,,
\end{equation} 
and 
\begin{equation}
    \label{equ:calphi}
    \cos\varphi'_{\rm m}=\frac{\cos\alpha\cos\rho'_{\rm m} - \cos(\alpha + \beta)}{\sin\alpha\sin\rho'_{\rm m}}
    \,.
\end{equation}

The position vector $\mathbf{r}'_{\rm m}$ (from the stellar center to the emission point) lies in the plane of a specific field line and shares the same $\varphi'_{\rm m}$ with $\mathbf{k}'_{\rm m}$. Its colatitude $\theta'_{\rm m}$ is related to $\rho'_{\rm m}$ by 
\begin{equation}
    \label{equ:theta-rho}
    \theta'_{\rm m}=\frac{1}{2}\arccos\left[\frac{\sqrt{\sin^4\rho'_{\rm m}-10\sin^2\rho'_{\rm m}+9}-\sin^2\rho'_{\rm m}}{3}\right]
\end{equation} 
\citep{Qiao+2001aap, Lee+2009apj}. The magnitude of $\mathbf{r}'_{\rm m}$, i.e., the emission height $h_{\rm em}$, depends on the field line configuration.

Rather than confining emission solely to the last open field lines, we introduce a footprint parameter $\eta$ to characterize the inner field lines. Here, $\eta = 1$ corresponds to the last open field lines, while smaller values of $\eta$ describe field lines progressively closer to the magnetic axis. The emission height in this case is 
\begin{equation}
    \label{equ:inneropen}
    h_{\rm em}=\frac{R_*}{\sin^2\left(\eta\arcsin\sqrt{\frac{R_*}{R_{\rm e}}}\right)}\sin^2\theta'_{\rm m}
\end{equation}
(see Appendix \ref{sec:field&trans} for derivation), where $R_{\rm e}$ is a function of $\varphi'_{\rm m}$ and $\alpha$: 
\begin{equation}
    \label{equ:re}
    R_{\rm e}=\frac{R_{\rm LC}}{\sin^2\theta_{\rm M} \sqrt{1-(\cos\alpha \cos\theta_{\rm M} - \sin\alpha \sin\theta_{\rm M} \cos\varphi'_{\rm m})^2}} \,,
\end{equation} 
$\theta_{\rm M}$ satisfies 
\begin{equation}
    \label{equ:thetaM}
    \begin{split}
        & 2\cos\theta_{\rm M} (\cos\theta_{\rm M} \sin\alpha + \cos\alpha \cos\varphi’_{\rm m} \sin\theta_{\rm M})^2\\
        & + \sin\theta_{\rm M} (\cos\alpha \cos\theta_{\rm M} \cos\varphi'_{\rm m} - \sin\alpha \sin\theta_{\rm M})(\cos\theta_{\rm M} \sin\alpha\\
        & + \cos\alpha \cos\varphi'_{\rm m} \sin\theta_{\rm M} + 3\cos\theta_{\rm M} \sin^2\theta_{\rm M} \sin^2\varphi'_{\rm m}) \\
        &=0
    \end{split}
\end{equation}
(see \citeauthor{Lee+2009apj} \citeyear{Lee+2009apj} for derivation), and $R_* = 10 \, \rm km$ is the neutron star radius.

Based on this static dipole modeling with a fixed impact angle, for a layer of field lines governed by a certain $\eta$ value, the intrinsic dependence of emission heights on intrinsic emission longitudes can be derived, as shown by the magenta U-shaped curve in the lower panel of Figure \ref{fig:arcurves}. Assuming the electric vector is in the plane of a field line, the intrinsic PA swing is calculated from spherical geometry \citep{Komesaroff1970Natur}: 
\begin{equation} \label{equ:paswing}
    \rm{PA} = \arctan \left( \frac{\sin \alpha \sin \phi_{\rm e}}{\sin \zeta \cos \alpha - \cos \zeta \sin \alpha \cos \phi_{\rm e}} \right) \,,
\end{equation}
where $\zeta = \alpha + \beta$, as shown in magenta in the upper panel of Figure \ref{fig:arcurves}.

The vectors $\mathbf{k}'_{\rm m}$ and $\mathbf{r}'_{\rm m}$ in spherical coordinates are thus derived for each intrinsic emission longitude. Transforming them to Cartesian coordinates and expressing them in $\mathcal{K}'$ as $\mathbf{k}'$ and $\mathbf{r}'$ (see Appendix \ref{sec:field&trans}), with the rotational velocity vector $\mathbf{v}_{\rm r} = \boldsymbol{\Omega} \times \mathbf{r}'$, we can apply the Lorentz transformation to $\mathbf{k}'$.

\subsubsection{A/R phase shifts}
\label{subsubsec:abe&ret}

First, the emission direction $\mathbf{k}'$ is expressed in $\mathcal{K}''$ as $\mathbf{k}''$ using the transformation matrix in Appendix \ref{sec:field&trans}. The Lorentz transformation
\begin{equation}
    \left\{
    \begin{split}
        & k_{\rm a, \, x}''=\frac{k_{\rm x}'' + \beta_{\rm r}}{1 + \beta_{\rm r}k_{\rm x}''}\\
        & k_{\rm a, \, y}''=\frac{k_{\rm y}''}{\gamma(1 + \beta_{\rm r}k_{\rm x}'')}\\
        & k_{\rm a, \, z}''=\frac{k_{\rm z}''}{\gamma(1 + \beta_{\rm r}k_{\rm x}'')}
    \end{split}
    \right.,
\end{equation}
where $\beta_{\rm r}=v_{\rm r}/c$ and $\gamma=1/\sqrt{1 - \beta_{\rm r}^2}$, is applied to $\mathbf{k}''$ to obtain the aberrated direction $\mathbf{k}_{\rm a}''$. This vector is then transformed back to $\mathcal{K}'$ to give $\mathbf{k}'_{\rm a}$. The aberrated emission longitude $\phi_{\rm a}$ is the angle $\mathbf{k}'_{\rm a}$ makes with the meridian plane, i.e., $\phi_{\rm a} = \arctan(k'_{\rm a,\, y}/k'_{\rm a,\, x})$. Because observed emission longitudes are defined in a left-handed system, we take the negative value of $\phi_{\rm a}$ as the observed emission longitude. The aberrated polarization direction is also determined. The aberrated PA can be calculated by substituting $\phi_{\rm e}$ with $\phi_{\rm a}$ and $\zeta$ with $\zeta_{\rm a} = \arccos k_{\rm a, \, z}'$ in Equation (\ref{equ:paswing}).

For the retardation effect, the propagation time difference between the stellar center and the emission point is $\Delta t = h_{\rm em} \cos \delta/c$, where $\delta$ is the angle between $\mathbf{r}'$ and $\mathbf{k}'_{\rm a}$. The corresponding phase advance is 
\begin{equation}
    \Delta\phi_{\rm r} = \frac{2\pi}{P}\Delta t = \frac{h_{\rm em}\cos\delta}{R_{\rm LC}} \,.
\end{equation}

\begin{figure}
    \centering
    \includegraphics[width=0.75\linewidth]{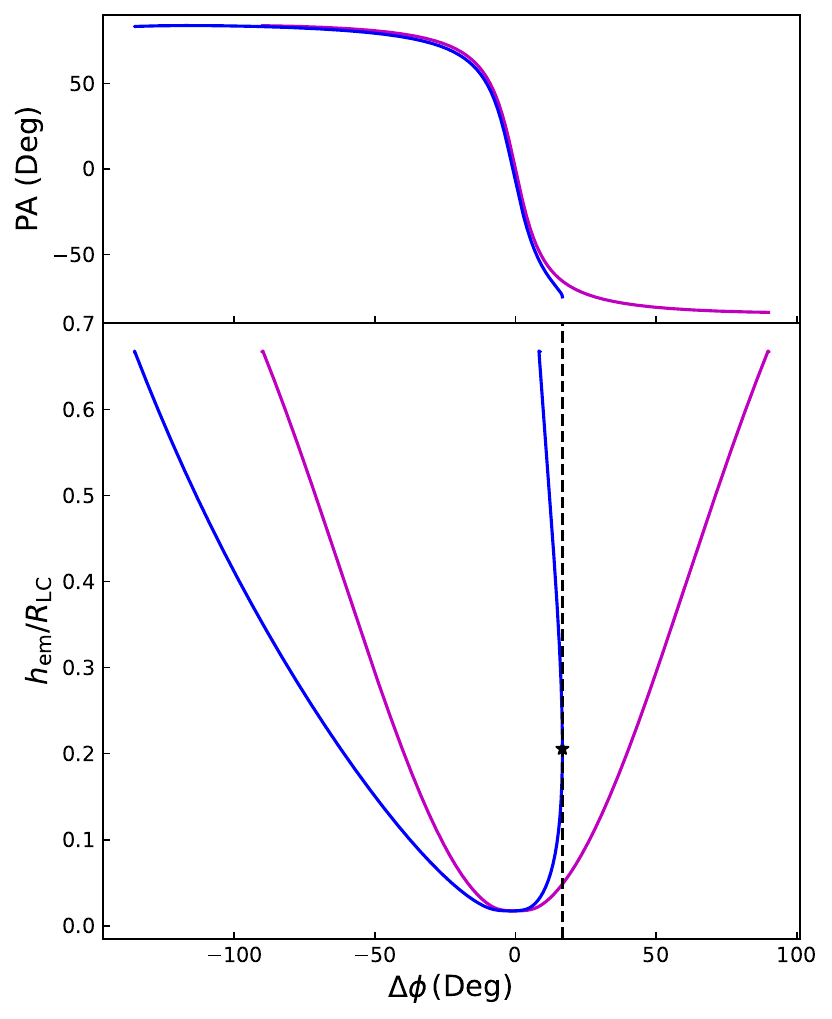}
    \caption{Example of the intrinsic and A/R-corrected PA swing (upper panel) and curves of emission height versus longitude (lower panels). Intrinsic curves are plotted in magenta; A/R-corrected ones are in blue. Note that the longitudes are shifted leftward after an inflection point (black star), and the PA swing corresponding to those longitudes is not plotted.}
    \label{fig:arcurves}
\end{figure}

The total A/R-shifted longitude $\phi_{\rm ar} = - \phi_{\rm a} - \Delta \phi_{\rm r}$ represents the observed longitude. Figure \ref{fig:arcurves} shows an example of the A/R-corrected PA swing and $h_{\rm em} - \Delta \phi$ curve. In the lower panel, the blue curve shows the A/R-corrected $h_{\rm em} - \Delta \phi$ relation, displaying a left-skewed ``U'' shape because higher altitudes yield larger phase shifts. Consequently, the A/R-corrected $h_{\rm em} - \Delta \phi$ curve may have an inflection point on its right side (black star in Figure \ref{fig:arcurves}). Beyond this inflection point, $\Delta \phi$ decreases, making the curve a double-valued function over a certain $\Delta \phi$ range. Moreover, after the inflection point, the A/R-corrected PA swing no longer follows the previous trend. This portion of the PA swing is therefore omitted in the upper panel. The remaining A/R-corrected PA swing is used to fit the observed PAs and derive the intrinsic emission geometry and emission longitudes.

\subsection{Constraint on the emission geometry}
\label{subsec:geo}

For simplicity, we assume that the observed emission from one pole originates from the field lines characterized by a single footprint parameter $\eta$, though $\eta$ may differ between the two poles. Hence, the A/R-corrected RVM for interpulse pulsars has six free parameters: $\alpha$, $\beta$, $\eta_{\rm M}$, $\eta_{\rm I}$, $\phi_0$, $\rm PA_0$, where $\eta_{\rm M}$ and $\eta_{\rm I}$ are the $\eta$ values for the main pulse and interpulse, respectively. Here, $\alpha$ and $\beta$ refer to the main pulse; the corresponding angles for the interpulse are derived from Equations (\ref{equ:alpha_trans}) and (\ref{equ:beta_trans}). 

\begin{deluxetable}{ccc}
\tablecaption{Priors on the A/R-corrected RVM parameters \label{tab:priors}}
\tablehead{
\colhead{Parameter} & \colhead{Prior Type} & \colhead{Range (Units)}
}
\startdata
$\alpha$ & Uniform & $0 \sim 180$ ($^\circ$) \\
$\beta$ & Uniform & $-45 \sim 45$ ($^\circ$) \\
$\eta_{\rm M}$ & Uniform & $0 \sim 1$ \\
$\eta_{\rm I}$ & Uniform & $0 \sim 1$ \\
$\phi_0$ & Uniform & $-90 \sim 90$ ($^\circ$) \\
$\rm PA_0$ & Uniform & $-90 \sim 90$ ($^\circ$) \\
\enddata
\end{deluxetable}

To constrain all six parameters, we use the \textsc{Python} library \textsc{Bilby}\footnote{\url{https://lscsoft.docs.ligo.org/bilby/}} \citep{bilby_paper} to perform Bayesian inference. The priors are listed in Table \ref{tab:priors}. Combining a Gaussian likelihood, we employ the \textsc{Dynesty} nested sampler \citep{Skilling2004aipc, skilling2006nested, Speagle2020mnras, Higson+2019sc} for sampling and obtain the posterior distributions.

An example of the posterior distributions is shown in Figure \ref{fig:corner}. The figures in Appendix \ref{sec:prof} present the A/R-corrected PA swing solutions (blue) and the corresponding intrinsic PA swings (magenta) from 200 samples within a $1\sigma$ confidence interval to visualize the fitting uncertainty. To ensure a clean fit, we excluded PAs with low linear polarization, those that deviated significantly from the overall trend, or those showing gradual transitions between polarization modes (plotted in gray). PAs consistent with the overall trend after a manual $90^\circ$ jump (indicating a different orthogonal polarization mode (OPM)) are shown in blue. These PAs are used for fitting after applying a $90^\circ$ jump. Median values and $1\sigma$ errors are reported in Table \ref{tab:res}.

Note that, for PSRs J1126$-$6054, J2047$+$5029, and J2208$+$4056, the inferred impact angles $\beta$ are remarkably large, and the result suggests a single-pole emission geometry, which does not align with our goal of comparing the emission beams from both poles. These pulsars are excluded from subsequent analyses.

\begin{figure}
    \centering
    \includegraphics[width=1.0\linewidth]{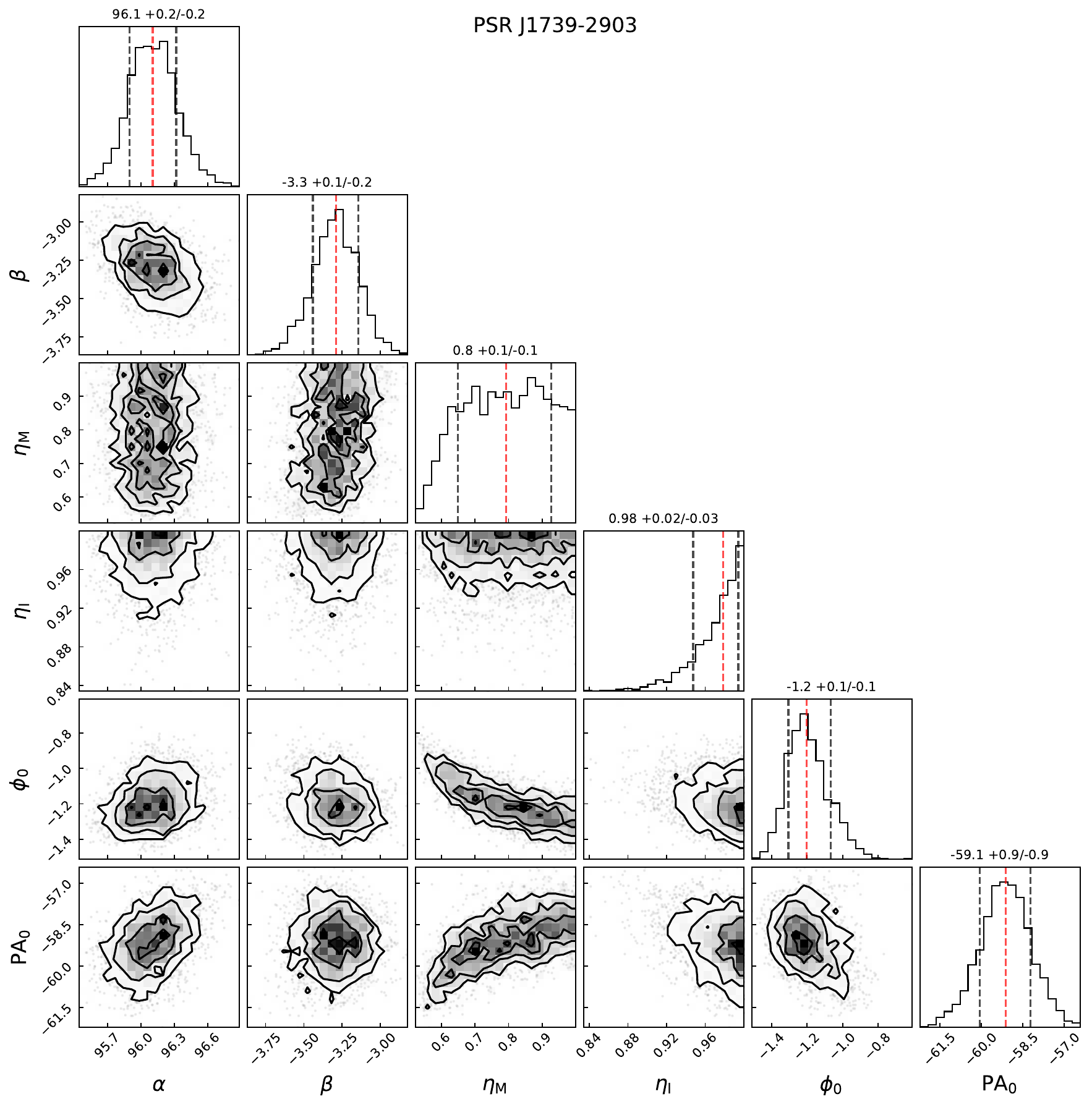}
    \caption{Example corner plot for PSR J1739$-$2903. One-dimensional marginal distributions are shown at the top of each column, in which black dashed lines indicate the $1\sigma$ quantiles and the red dashed lines the median values. Correlations between parameters are displayed in other panels.}
    \label{fig:corner}
\end{figure}

\begin{deluxetable*}{cccccccc}
\tablewidth{\textwidth}
\tablecaption{Intrinsic emission geometry obtained with the A/R-corrected RVM. \label{tab:res}}
\tablehead{
\colhead{PSRJ} & \colhead{$\alpha_{\rm M}$} & \colhead{$\beta_{\rm M}$} & \colhead{$\eta_{\rm M}$} & \colhead{$\phi_0$} & \colhead{$\alpha_{\rm I}$} & \colhead{$\beta_{\rm I}$} & \colhead{$\eta_{\rm I}$} \\ \colhead{} & \colhead{(deg)} & \colhead{(deg)} & \colhead{} & \colhead{(deg)} & \colhead{(deg)} & \colhead{(deg)} & \colhead{}
}
\startdata
0908$-$4913 & $83.773^{+0.008}_{-0.007}$ & $7.52^{+0.02}_{-0.02}$ & $0.9990^{+0.0007}_{-0.0016}$ & $4.434^{+0.006}_{-0.006}$ & $96.227^{+0.008}_{-0.007}$ & $-4.93^{+0.03}_{-0.02}$ & $0.99^{+0.01}_{-0.02}$ \\
1126$-$6054$\dagger$ & $158^{+11}_{-9}$ & $-1.7^{+0.9}_{-0.8}$ & $0.6^{+0.3}_{-0.3}$ & $-0.3^{+0.2}_{-0.2}$ & $22^{+11}_{-9}$ & $134.3^{+22}_{-18}$ & $0.8^{+0.2}_{-0.2}$ \\
1413$-$6307 & $98.2^{+0.3}_{-0.3}$ & $-2.53^{+0.05}_{-0.05}$ & $0.93^{+0.05}_{-0.07}$ & $-3.08^{+0.04}_{-0.04}$ & $81.8^{+1}_{-1}$ & $13.9^{+0.6}_{-0.6}$ & $0.98^{+0.01}_{-0.02}$ \\
1549$-$4848 & $87.5^{+0.1}_{-0.1}$ & $2.9^{+0.3}_{-0.3}$ & $0.26^{+0.02}_{-0.01}$ & $2.8^{+0.2}_{-0.2}$ & $92.5^{+0.1}_{-0.1}$ & $-2.1^{+0.4}_{-0.4}$ & $0.8^{+0.2}_{-0.2}$ \\
1611$-$5209 & $96.6^{+1.2}_{-0.8}$ & $-6.0^{+0.3}_{-0.4}$ & $0.7^{+0.2}_{-0.1}$ & $-0.3^{+0.7}_{-0.5}$ & $83.4^{+1.2}_{-0.8}$ & $7^{+2}_{-2}$ & $0.8^{+0.1}_{-0.2}$ \\
1739$-$2903 & $96.1^{+0.2}_{-0.2}$ & $-3.3^{+0.1}_{-0.2}$ & $0.8^{+0.1}_{-0.1}$ & $-1.2^{+0.1}_{-0.1}$ & $83.9^{+0.2}_{-0.2}$ & $8.9^{+0.4}_{-0.4}$ & $0.98^{+0.02}_{-0.03}$ \\
1755$-$0903 & $70.4^{+0.8}_{-1.0}$ & $5.7^{+0.2}_{-0.2}$ & $0.8^{+0.1}_{-0.2}$ & $3.1^{+0.3}_{-0.3}$ & $109.6^{+0.8}_{-1.0}$ & $-34^{+2}_{-2}$ & $0.920^{+0.004}_{-0.003}$ \\
1909$+$0749 & $98.0^{+0.7}_{-0.6}$ & $-2^{+1}_{-1}$ & $0.6^{+0.2}_{-0.3}$ & $10^{+1}_{-1}$ & $82.0^{+0.7}_{-0.6}$ & $14^{+2}_{-2}$ & $0.88^{+0.08}_{-0.09}$ \\
1913$+$0832 & $95.1^{+0.1}_{-0.1}$ & $-0.09^{+0.06}_{-0.11}$ & $0.5^{+0.3}_{-0.3}$ & $22.1^{+0.2}_{-0.2}$ & $84.9^{+0.1}_{-0.1}$ & $10.1^{+0.2}_{-0.2}$ & $0.78^{+0.03}_{-0.02}$ \\
2047$+$5029$\dagger$ & $161^{+8}_{-8}$ & $-0.8^{+0.3}_{-0.3}$ & $0.6^{+0.3}_{-0.3}$ & $-1.5^{+0.2}_{-0.2}$ & $19^{+17}_{-24}$ & $141^{+16}_{-16}$ & $0.6^{+0.2}_{-0.2}$ \\
2208$+$4056$\dagger$ & $151^{+2}_{-2}$ & $-6.8^{+0.6}_{-0.6}$ & $0.8^{+0.1}_{-0.2}$ & $1.3^{+0.2}_{-0.2}$ & $29^{+2}_{-2}$ & $115^{+4}_{-4}$ & $0.87^{+0.09}_{-0.14}$ \\
\enddata
\tablecomments{$\alpha$ and $\beta$ with subscript ``M'' denote the inclination and impact angles for the main pulse; those with ``I'' denote the interpulse. $\phi_0$ is the fiducial pulse longitude. $\eta_{\rm M}$ and $\eta_{\rm I}$ are the footprint parameters for the main pulse and interpulse, respectively. \\
$\dagger$ Discarded}
\end{deluxetable*}

\subsection{Intrinsic emission longitude}
\label{subsec:emregion}

After numerically fitting the observed PA swing with the A/R-corrected RVM, a one-to-one correspondence between the observed and the intrinsic emission longitudes is established. To find the intrinsic emission longitudes, we first determine the observed pulse windows.

\begin{figure}
    \centering
    \includegraphics[width=\linewidth]{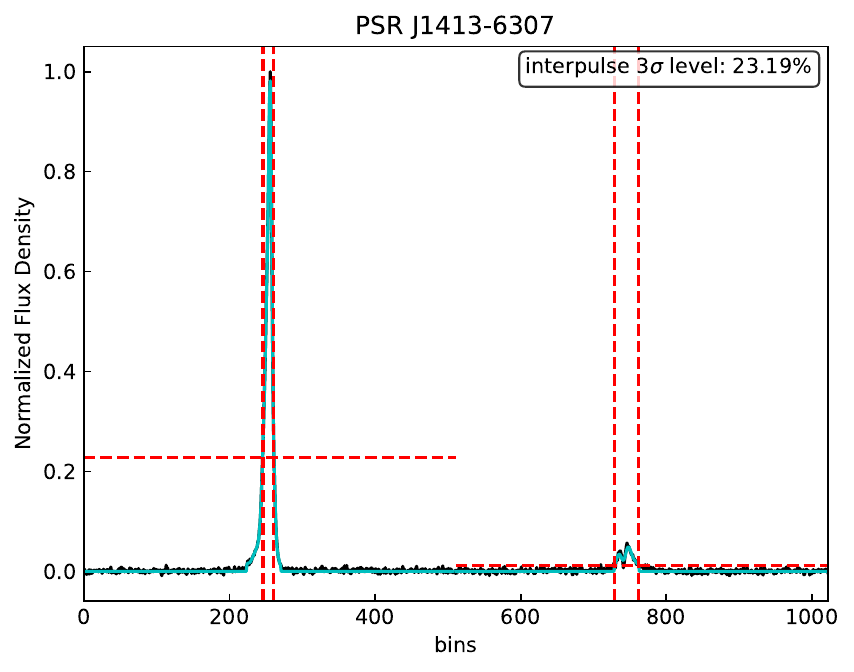}
    \caption{Example of a GP-smoothed profile and the determination of the pulse window. The two horizontal red dashed lines indicate the level (as a percentage of peak intensity) at which the pulse windows are measured. Vertical red dashed lines delineate the window boundaries.}
    \label{fig:winlevel}
\end{figure}

We use a Gaussian process (GP), implemented with the \textsc{George}\footnote{\url{https://george.readthedocs.io}} library \citep{Ambikasaran+2015itpm}, to separate the noise and signal components in the pulse profiles, yielding a smooth, noiseless profile. This enables us to measure the interpulse window at the lowest feasible level in the presence of noise \citep{Brook+2019mnras, Johnston+2019amnras, Posselt+2021mnras}. An example of the GP-smoothed profile is shown in Figure \ref{fig:winlevel}.

To ensure a fair comparison between the main pulse and the interpulse, their pulse windows must be defined at the same percentage level of their respective peak intensities. We first find the $3\sigma$ level for the interpulse. The corresponding percentage of this level relative to the interpulse peak is then applied to both the main pulse and the interpulse to define their windows. For example, in Figure \ref{fig:winlevel}, the $3\sigma$ level lies at $23.19\%$ of the interpulse peak intensity. The pulse windows of both the main pulse and interpulse are therefore determined at $23.19\%$ of their respective peaks (indicated with red horizontal dashed lines), as delineated with red vertical dashed lines.

For pulsars with interpulse emission much weaker than the main pulse, a higher percentage level may be used to determine the pulse window, potentially causing some of the main pulse components to be missed and therefore leading to a biased measurement for the beam parameters. The percentage levels applied for PSRs J1413-6307, J1549-4848, J1611-5209, and J1755-0903 exceed $20\%$, reaching even $60\%$ for PSR J1611-5209. However, except for PSR J1413-6307, no other significant components are present below the respective percentage levels applied for selected pulsars. Therefore, measurements and comparisons of the emission beam parameters at their two poles should be reliable. The leading edge of the main pulse of PSR J1413-6307 contains a component with an intensity comparable to that of the interpulse, which has not been included within the main pulse window. However, due to insufficient signal-to-noise ratio (S/N), we are unable to incorporate this component into the pulse window for more reliable comparisons.

\begin{figure*}
    \centering
    \includegraphics[width=0.7\textwidth]{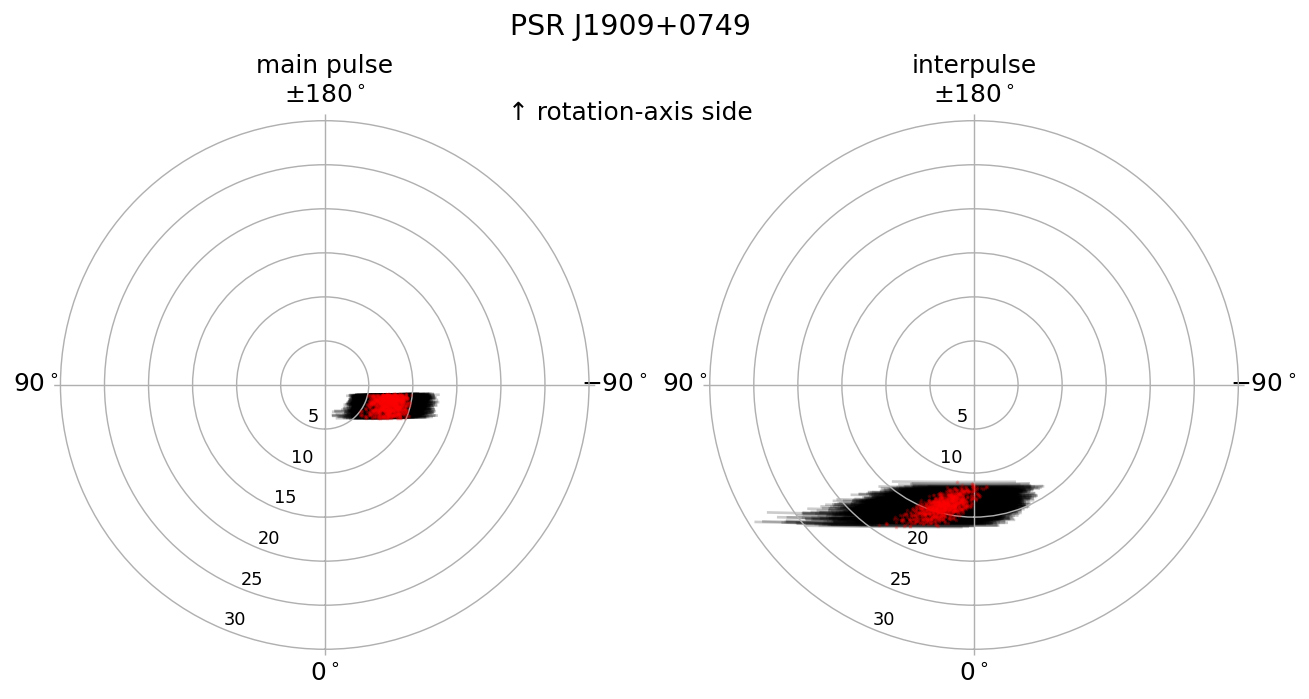}
    \caption{Possible intrinsic emission region mapping for PSR J1909$+$0749 as an example. The left and right panels show maps for the main pulse and interpulse, respectively. The upper hemisphere is the rotation-axis side; the lower hemisphere is the equatorial side (see Section \ref{sec:data} for definitions). Black solid lines are the possible intrinsic emission regions derived from posterior samples within the $1\sigma$ confidence interval. Red stars mark the possible intrinsic emission locations corresponding to the pulse profile peak derived from posterior samples within the $1\sigma$ confidence interval.}
    \label{fig:map}
\end{figure*}

Intrinsic emission longitudes are then solved for the observed emission longitudes within the determined pulse windows. However, as shown in Section \ref{subsubsec:abe&ret}, the longitudes of the A/R-corrected RVM reach a maximum at an inflection point (black dashed line in Figure \ref{fig:arcurves}). Smaller $\eta$ values correspond to higher emission altitudes and a more leftward inflection point. For sufficiently small $\eta$, the inflection point may fall left of the right boundary of the pulse window, leaving some observed pulse longitudes without intrinsic solutions. For such posterior samples, we step $\eta_{\rm supp}$ from 1 down to the problematic value in decrements of 0.02 to identify the range of $\eta$ values that yield solutions for the remaining observed emission longitudes. Possible intrinsic solutions for the remaining observed emission longitudes are solved with these $\eta_{\rm supp}$ values.

Additionally, the A/R-corrected RVM provides phase-resolved intrinsic emission heights. For a given emission geometry, the larger $\eta$ values yield lower emission heights. When the LOS passes very close to the magnetic axis (small $\beta$) and the $\eta$ value is near 1, the emission heights can become lower than the stellar radius. We discard posterior samples in this case. The possible ranges of intrinsic emission longitudes are derived from the remaining samples.

\section{Comparing the two-pole emission regions}
\label{sec:comp}

Using the possible intrinsic emission longitudes derived in Section \ref{sec:meth}, we calculate the intrinsic emission radii $\rho$ and magnetic azimuth angles $\varphi$ from Equations (\ref{equ:calrho}) and (\ref{equ:calphi}). This allows us to map the possible intrinsic emission regions for each magnetic pole, as shown in Figure \ref{fig:map} and the top panels of the figures in Appendix \ref{sec:comp}. From these maps, we identify the locations of active emission regions for both poles of each pulsar, summarized in Table \ref{tab:comp}. This information enables us to investigate whether the active emission regions at the two poles are correlated or randomly distributed. The possible emission radii and azimuth angles also allow a quantitative comparison of the emission beams from the two magnetic poles under the conal beam and fan beam models.

\subsection{Comparison under the conal beam model}
\label{subsec:conal}

In the conal beam model, the radio emission beam is assumed to be circular. Its angular extent, characterized by the emission radius $\rho$, reflects the size of the active emission region. We therefore use the maximum emission radius of the intrinsic emission region as an indicator for the overall beam scale, acknowledging that the emission regions are not always symmetric about the meridian plane, as shown in Figure \ref{fig:map} and the top panels of the figures in Appendix \ref{sec:comp}.

From the possible intrinsic emission regions, we extract the distributions of maximum emission radii for the main pulse ($\rho_{\rm M}$) and interpulse ($\rho_{\rm I}$). Joint distributions of $\rho_{\rm M}$ and $\rho_{\rm I}$ are shown in the left panel of Figure \ref{fig:dist} and the middle-left panels of the figures in Appendix \ref{sec:comp}. Black dots represent cases where all intrinsic emission longitudes are obtained with a single $\eta$ value; red dots indicate that some observed emission longitudes require solutions given by supplementary $\eta_{\rm supp}$ values (see Section \ref{subsec:emregion}). A red dashed line indicating $\rho_{\rm M} = \rho_{\rm I}$ is plotted in each panel. Similarity in beam radii is suggested if this line intersects the joint distribution.

\begin{figure*}[htp]
    \centering
    \includegraphics[width=0.7\textwidth]{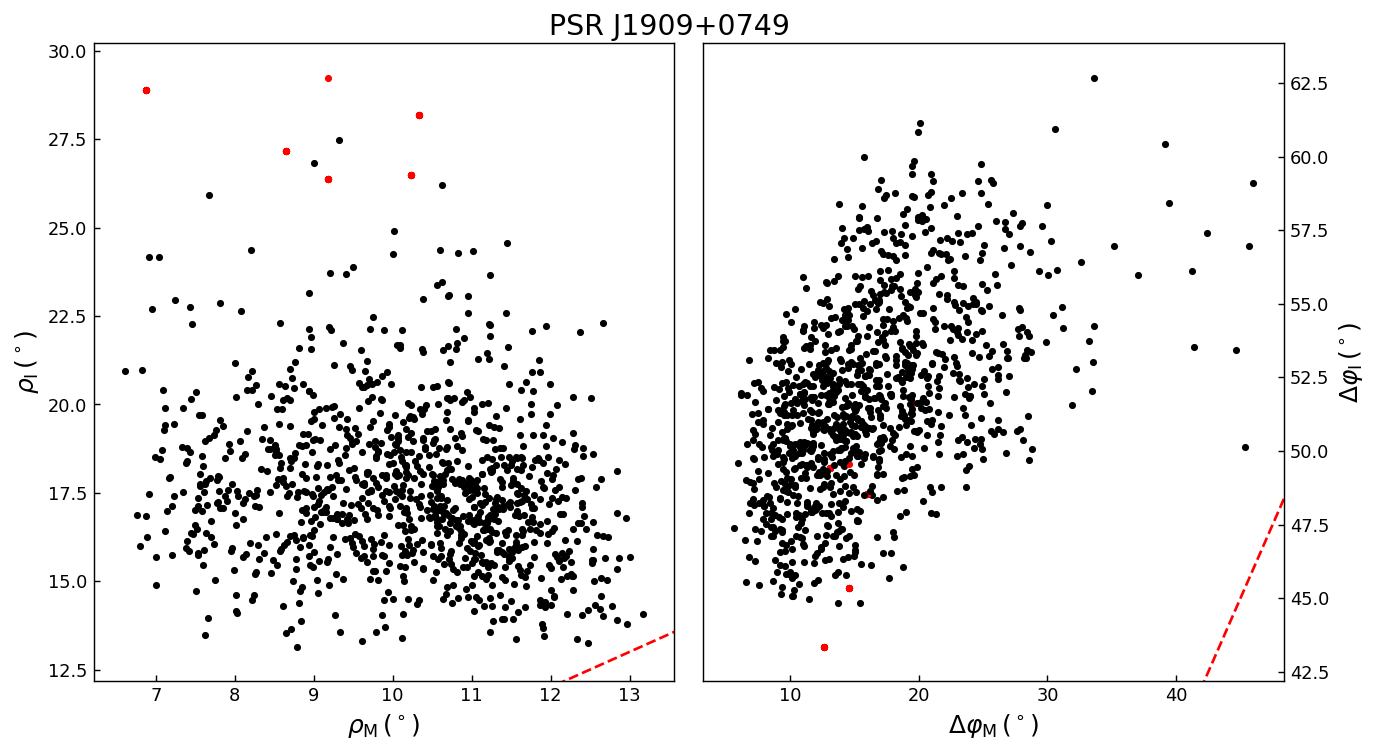}
    \caption{Example joint distribution of the beam radii, $\rho_{\rm M}$ versus $\rho_{\rm I}$ (left), and magnetic azimuth width, $\Delta \varphi_{\rm M}$ versus $\Delta \varphi_{\rm I}$ (right), for PSR J1909$+$0749. Distributions are derived from samples within the $1\sigma$ confidence interval of the emission geometry. Horizontal axes correspond to the main pulse, vertical axes to the interpulse. Red dashed lines indicate equality.}
    \label{fig:dist}
\end{figure*}

The results, summarized in the second column in Table \ref{tab:comp}, indicate that the emission beam radii of both poles may be similar only for PSRs J1549$-$4848 and J1611$-$5209.

\subsection{Comparison under the fan beam model}
\label{subsec:fanbeam}

The fan beam model assumes that the radio emission is broadband and forms a radially extended beam. Its lateral extent is characterized by the magnetic azimuth width $\Delta \varphi$, while the radial intensity profile follows the relation given by \citetalias{Wang+2014apj}: 
\begin{equation} \label{equ:I-rho_relation}
    I_{\rm peak}=I_0(P, \dot{P}, \alpha)\rho^{2q-6}_{\rm peak} \,,
\end{equation}
where $I_{\rm peak}$ is the peak intensity of the observed pulse profile, $\rho_{\rm peak}$ is the corresponding emission radius, $q$ is the radial intensity damping index, and $I_0$ represents the maximum intensity of the radio emission beam, which depends on the rotation period $P$, its derivative $\dot{P}$ and $\alpha$. Based on the parameter space of this model, we can compare both the magnetic azimuth widths and intensities of the emission beams from the two magnetic poles.

\subsubsection{Comparison of magnetic azimuth widths}
\label{subsubsec:deltaphi}

From the possible intrinsic emission regions, we extract the distributions of magnetic azimuth widths for the main pulse ($\Delta \varphi_{\rm M}$) and interpulse ($\Delta \varphi_{\rm I}$). Joint distributions are presented in the right panel of Figure \ref{fig:dist} and the middle-right panels of the figures in Appendix \ref{sec:comp}. Black dots represent cases solved with a single $\eta$ value; red dots indicate that supplementary $\eta_{\rm supp}$ values are needed (see Section \ref{subsec:emregion}). The red dashed line represents $\Delta \varphi_{\rm M} = \Delta \varphi_{\rm I}$; an intersection with the joint distribution suggests similar magnetic azimuth widths.

As summarized in the third column of Table \ref{tab:comp}, no pulsar shows potential similarity in the magnetic azimuth widths of its two poles within the explored parameter space.

\begin{figure}
    \centering
    \includegraphics[width=0.4\textwidth]{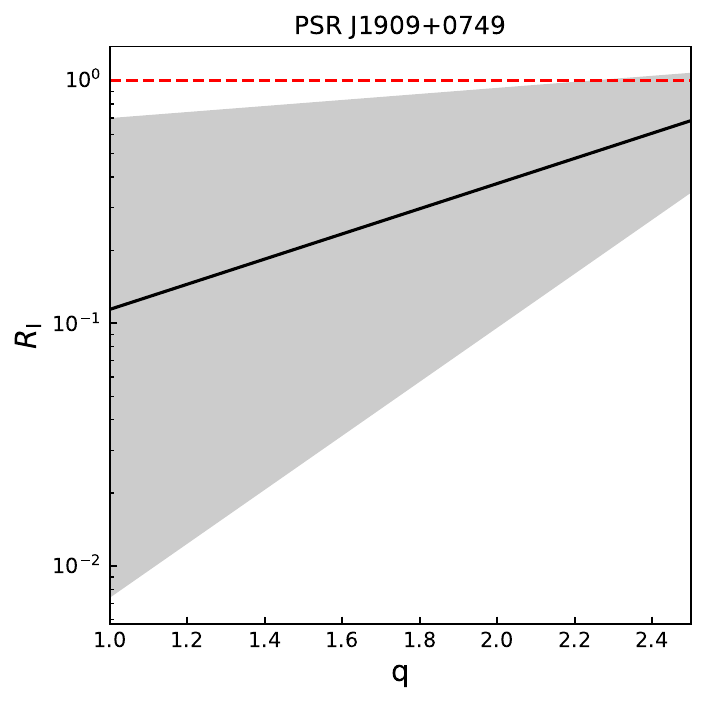}
    \caption{Example $R_{\rm I}-q$ relation for PSR J1909$+$0749. The shaded area represents the uncertainty from the $1\sigma$ confidence interval of the geometry parameters. The red horizontal dashed line indicates $R_{\rm I}=1$.}
    \label{fig:R_I-q}
\end{figure}

\subsubsection{Comparison of beam intensities}
\label{subsubsec:beamintensity}

\begin{deluxetable*}{cccccc}
\tablecaption{Comparison results and locations of active emission regions. The columns labeled $\rho_{\rm M} = \rho_{\rm I}$?, $\Delta \varphi_{\rm M} = \Delta \varphi_{\rm I}$?, and $R_{\rm I} = 1$? indicate the similarity tests for the emission beam radii, magnetic azimuthal widths, and emission intensities, respectively. A ``\textbf{Y}'' denotes similarity, while an ``N'' denotes dissimilarity. The final two columns specify the active emission regions for the main pulse and interpulse: ``L'' indicates the leading side, ``T'' indicates the trailing side, and ``S'' indicates symmetry with respect to the meridian plane. \label{tab:comp}}

\tablewidth{\textwidth}

\tablehead{
PSRJ & $\rho_{\rm M} = \rho_{\rm I}$? & $\Delta \varphi_{\rm M} = \Delta \varphi_{\rm I}$? & $R_{\rm I} = 1$? & \multicolumn{2}{c}{Active Regions} \\
 & (\textbf{Y}/N) & (\textbf{Y}/N) & (\textbf{Y}/N) & Main Pulse & Interpulse
}
\startdata
0908$-$4913 & N & N & N & T & T \\
1413$-$6307 & N & N & \textbf{Y} & L & L \\
1549$-$4848 & \textbf{Y} & N & \textbf{Y} & L & T \\
1611$-$5209 & \textbf{Y} & N & N & L & S \\
1739$-$2903 & N & N & \textbf{Y} & L & L \\
1755$-$0903 & N & N & \textbf{Y} & S & L \\
1909$+$0749 & N & N & \textbf{Y} & T & L \\
1913$+$0832 & N & N & \textbf{Y} & T & T \\
\enddata
\end{deluxetable*}

To compare the emission beam intensities, we calculate the ratio of the maximum beam intensities from the two poles as
\begin{equation}
    R_{\rm I}=\frac{I_{0, \rm MP}}{I_{0, \rm IP}}=\frac{I_{\rm peak, \rm MP}}{I_{\rm peak, \rm IP}}\left(\frac{\rho_{\rm peak, \rm MP}}{\rho_{\rm peak, \rm IP}}\right)^{-2q+6}
    \,.
\end{equation}
Here, the damping index $q$, characterizing the radial decay of emission intensity, is constrained to the range $(1, \, 2.5)$ based on fits to 64 pulsars by \citetalias{Wang+2014apj}. Within this range of parameter $q$, we examine potential similarity for the emission intensities from the two poles.

The relations between $R_{\rm I}$ and $q$ are shown in Figure \ref{fig:R_I-q} and the lower panels of the figures in Appendix \ref{sec:comp}. Shaded regions represent uncertainties from the fitted emission geometry. The red dashed line at $R_{\rm I}=1$ indicates equal beam intensity. Similarity is suggested if this line intersects the uncertainty region.

The results, shown in the fourth column of Table \ref{tab:res}, indicate that the beam intensities from the two poles of PSRs J1413$-$6307, J1549$-$4848, J1739$-$2903, J1755$-$0903, J1909$+$0749, and J1913$+$0832 are potentially similar within a specific parameter space.

\section{Individual pulsars}
\label{sec:psrs}

\subsection*{PSR J0908$-$4913}

This pulsar has a relatively high S/N. The $3\sigma$ level of the interpulse is located at $1.13\%$ of its peak intensity, and the main pulse window is also determined at this percentage level relative to the main pulse peak. Linear polarization is significant across its whole on-pulse region, producing an extended and regular PA swing. Consequently, the emission geometry is precisely constrained. The inclination and impact angles derived here with the A/R-corrected RVM are $83.773^\circ{}^{+0.008}_{-0.007}$ and $7.52^\circ{}^{+0.02}_{-0.02}$, consistent with those reported by \citetalias{Johnston+2019bmnras}. The emission heights for possible intrinsic emission regions of this pulsar range from $87 \, \rm km$ ($0.017 \, R_{\rm LC}$) to $196 \, \rm km$ ($0.04 \, R_{\rm LC}$) for the main pulse, and from $37 \, \rm km$ ($0.007 \, R_{\rm LC}$) to $231 \, \rm km$ ($0.045 \, R_{\rm LC}$) for the interpulse. Considering that upper boundaries correspond to the edge of the emission region—where the emission height was typically measured previously—these values fall within the ranges given by \citetalias{Johnston+2019bmnras}. The emission beams from the two poles show no similarity in any of the three beam parameters. The maps of intrinsic emission regions suggest that active emission regions are likely located on the trailing side of both magnetic poles.

\subsection*{PSR J1126$-$6054}

The PA swing of the leading linear polarization component of the main pulse is disordered, while that of the trailing component aligns with the RVM. Although the fractional linear polarization of the interpulse is high, its PA swing is not very regular, possibly due to the limited S/N. \cite{Maciesiak+2011mnras} identified this pulsar as an orthogonal rotator according to the main pulse-interpulse separation and PA swing. In this work, a single-pole emission geometry is obtained, but it is not robust given the interpulse PA swing. Further higher-sensitivity observations are needed to reveal a more reliable interpulse PA swing and emission geometry.

\subsection*{PSR J1413$-$6307}

The $3\sigma$ level of the interpulse is located at $23.19\%$ of its peak intensity, and the main pulse window is also determined at this percentage level relative to the main pulse peak. A gradual OPM jump occurs between the leading (with circular polarization) and trailing parts of the main pulse PA swing. The interpulse PA swing is not perfectly regular but maintains an overall RVM-like trend. The emission geometry, not previously reported, is obtained here as $\alpha = 98.2^\circ{}^{+0.3}_{-0.3}$ and $\beta = -2.53^\circ{}^{+0.05}_{-0.05}$, confirming an orthogonal rotator, as classified by \cite{Maciesiak+2011mnras}. Possible emission heights of this pulsar range from $35 \, \rm km$ ($0.002 \, R_{\rm LC}$) to $147 \, \rm km$ ($0.008 \, R_{\rm LC}$) for the main pulse, and from $981 \, \rm km$ ($0.05 \, R_{\rm LC}$) to $1330 \, \rm km$ ($0.07 \, R_{\rm LC}$) for the interpulse. Only the beam intensities of both poles show potential similarity. Intrinsic emission region maps of this pulsar suggest that active emission regions are likely located on the leading side of both magnetic poles.

\subsection*{PSR J1549$-$4848}

The $3\sigma$ level of the interpulse is located at $20.05\%$ of its peak intensity, and the main pulse window is also determined at this percentage level relative to the main pulse peak. This pulsar displays moderate linear polarization. While the overall main pulse PA swing follows an S-shaped curve, both its leading and trailing portions show deviations from the RVM. These segments were also excluded in the analysis by \citetalias{Johnston+2019bmnras}. The remaining main pulse PA swing exhibits weak negative circular polarization. The trailing part of the interpulse PA swing exhibits an OPM different from the leading part. Our A/R-corrected RVM fit gives $\alpha = 87.5^\circ{}^{+0.1}_{-0.1}$ and $\beta = 2.9^\circ{}^{+0.3}_{-0.3}$, consistent with \citetalias{Johnston+2019bmnras}. However, the derived $\eta_{\rm M}$ is small, so intrinsic solutions are only obtained for part of the main pulse window. In contrast, $\eta_{\rm I}$ ranges from 0.6 to 1.0, providing full coverage across the observed interpulse window. Possible emission heights of this pulsar range from $345 \, \rm km$ ($0.025 \, R_{\rm LC}$) to $2659 \, \rm km$ ($0.2 \, R_{\rm LC}$) for the main pulse, and from $13 \, \rm km$ ($0.001 \, R_{\rm LC}$) to $586 \, \rm km$ ($0.04 \, R_{\rm LC}$) for the interpulse. Because the A/R-corrected emission longitudes from all $\eta_{\rm M}$ samples do not cover the full main pulse window, supplementary $\eta_{\rm supp}$ values are used to obtain full coverage, and the samples in $\rho_{\rm M} - \rho_{\rm I}$ and $\Delta \varphi_{\rm M} - \Delta \varphi_{\rm I}$ are marked in red. The beam radii and intensities of the two poles are potentially similar. However, the active emission regions at its two poles appear on opposite sides: leading for the main pulse, trailing for the interpulse.

\subsection*{PSR J1611$-$5209}

Given a weak interpulse emission, the $3\sigma$ level of the interpulse is located at $60.24\%$ of its peak intensity, and the main pulse window is also determined at this percentage level relative to the main pulse peak. The fractional linear polarization of this pulsar is moderate, limiting the number of reliable PAs. The interpulse PA swing is not perfectly regular but shows an overall RVM trend. This pulsar was identified as having a double-pole geometry by \cite{Maciesiak+2011mnras}. Our fit gives $\alpha = 96.6^\circ{}^{+1.2}_{-0.8}$ and $\beta = -6.0^\circ{}^{+0.3}_{-0.4}$, confirming an orthogonal rotator. Possible emission heights of this pulsar range from $105 \, \rm km$ ($0.012 \, R_{\rm LC}$) to $378 \, \rm km$ ($0.04 \, R_{\rm LC}$) for the main pulse, and from $90 \, \rm km$ ($0.01 \, R_{\rm LC}$) to $644 \, \rm km$ ($0.07 \, R_{\rm LC}$) for the interpulse. Only the beam radii of the two poles of this pulsar are potentially similar. Intrinsic emission region maps of this pulsar suggest that the main pulse active region is on the leading side, while the interpulse region is likely symmetric about the meridian plane.

\subsection*{PSR J1739$-$2903}

The $3\sigma$ level of the interpulse is located at $7.85\%$ of its peak intensity, and the main pulse window is also determined at this percentage level relative to the main pulse peak. This pulsar has relatively low fractional linear polarization. The PAs of the first two linear polarization components of the main pulse exhibit a different polarization mode from the last component, which is accompanied by positive circular polarization. Our fit gives $\alpha = 96.1^\circ{}^{+0.2}_{-0.2}$ and $\beta = -3.3^\circ{}^{+0.1}_{-0.2}$, consistent with \citetalias{Johnston+2019bmnras}. Possible emission heights of this pulsar range from $54 \, \rm km$ ($0.0035 \, R_{\rm LC}$) to $417 \, \rm km$ ($0.027 \, R_{\rm LC}$) for the main pulse, and from $341 \, \rm km$ ($0.02 \, R_{\rm LC}$) to $674 \, \rm km$ ($0.044 \, R_{\rm LC}$) for the interpulse, consistent with previous estimates by \citetalias{Johnston+2019bmnras}. Only the beam intensities of both poles show potential similarity. Intrinsic emission region maps of this pulsar indicate that active emission regions are likely located on the leading side of both magnetic poles.

\subsection*{PSR J1755$-$0903}

Given the limited S/N, the $3\sigma$ level of the interpulse is located at $51.04\%$ of its peak intensity, and the main pulse window is also determined at this percentage level relative to the main pulse peak. The sign of circular polarization flips near the profile center. Both leading and trailing parts of the main pulse PA swing correspond to positive circular polarization, with a jump of less than $90^\circ$ between them \citep{Sun+2025apj}. Between the leading and trailing PA components of the main pulse, a gradual OPM transition occurs, accompanied by a sign reversal of circular polarization. Only the trailing part of the main pulse PA swing is used here. Our A/R-corrected RVM fit yields $\alpha = 70.4^\circ{}^{+0.8}_{-1.0}$ and $\beta = 5.7^\circ{}^{+0.2}_{-0.2}$, significantly different from the results reported by \cite{Serylak+2021MNRAS} and \cite{Sun+2025apj}. This discrepancy may be due to the large interpulse impact angle (and hence large emission height). Possible emission heights of this pulsar range from $72 \, \rm km$ ($0.008 \, R_{\rm LC}$) to $210 \, \rm km$ ($0.023 \, R_{\rm LC}$) for the main pulse, and from $2388 \, \rm km$ ($0.26 \, R_{\rm LC}$) to $7439 \, \rm km$ ($0.8 \, R_{\rm LC}$) for the interpulse. The substantial interpulse emission height distorts the A/R-corrected RVM. The emission beams of the two poles could be similar only in beam intensity. Intrinsic emission region maps of this pulsar suggest that the main pulse active emission region is likely symmetric about the meridian plane, while the interpulse active region is on the leading side.

\subsection*{PSR J1909$+$0749}

The $3\sigma$ level of the interpulse is located at $17.78\%$ of its peak intensity, and the main pulse window is also determined at this percentage level relative to the main pulse peak. The main pulse fractional linear polarization is moderate, while the interpulse shows very high fractional linear polarization. \cite{Wang+2023raa} and \cite{Sun+2025apj} applied a $180^\circ$ jump to the main pulse PA swing to place it above the interpulse PA swing, but with this treatment, our A/R-corrected RVM fit failed to converge. Here, we keep the main pulse PA swing below the interpulse and perform the fit. We obtain $\alpha = 98.0^\circ{}^{+0.7}_{-0.6}$ and $\beta = -2^\circ{}^{+1}_{-1}$, also consistent with \citet[][in which they reported the geometry for the interpulse]{Wang+2023raa} and \cite{Sun+2025apj}. Possible emission heights of this pulsar range from $31 \, \rm km$ ($0.003 \, R_{\rm LC}$) to $871 \, \rm km$ ($0.08 \, R_{\rm LC}$) for the main pulse, and from $449 \, \rm km$ ($0.04 \, R_{\rm LC}$) to $2226 \, \rm km$ ($0.2 \, R_{\rm LC}$) for the interpulse. Intrinsic emission longitude solutions given by several $\eta_{\rm I}$ samples do not cover the entire interpulse window. $\rho_{\rm M} - \rho_{\rm I}$ and $\Delta \varphi_{\rm M} - \Delta \varphi_{\rm I}$ samples marked in red indicate that a part of the interpulse window needs solutions given by supplementary $\eta_{\rm supp}$ values. The emission beams from both poles are potentially similar only in emission intensity. Intrinsic emission region maps of this pulsar indicate that active emission regions at both poles are on opposite sides: trailing for the main pulse, leading for the interpulse.

\subsection*{PSR J1913$+$0832}

The $3\sigma$ level of the interpulse is located at $8.14\%$ of its peak intensity, and the main pulse window is also determined at this percentage level relative to the main pulse peak. This pulsar has relatively high fractional linear polarization. Main pulse PAs are nearly constant across the window, while the interpulse shows an S-shaped PA swing. \cite{Wang+2023raa} and \cite{Sun+2025apj} presented two different treatments for the PA swing, yielding entirely different results. \cite{Wang+2023raa} applied a $-180^\circ$ jump to the main pulse PA swing, whereas \cite{Sun+2025apj} did not. Following \cite{Sun+2025apj} leads to a gradually decreasing fitted main pulse PA swing that does not match the flat observed PAs. The treatment by \cite{Wang+2023raa} produces a flat main pulse PA swing, which appears more realistic. Adopting the latter, we obtain $\alpha = 95.1^\circ{}^{+0.1}_{-0.1}$ and $\beta = -0.09^\circ{}^{+0.06}_{-0.11}$, consistent with \citet[][in which they reported the geometry for the interpulse]{Wang+2023raa}. Possible emission heights of this pulsar range from $10 \, \rm km$ ($0.0016 \, R_{\rm LC}$) to $1737 \, \rm km$ ($0.27 \, R_{\rm LC}$) for the main pulse, and from $308 \, \rm km$ ($0.05 \, R_{\rm LC}$) to $1211 \, \rm km$ ($0.2 \, R_{\rm LC}$) for the interpulse. No $\eta_{\rm I}$ sample yields intrinsic emission longitude solutions across the entire interpulse window. However, only a few samples have possible $\eta_{\rm supp}$ values to solve for the remaining interpulse window. $\rho_{\rm M} - \rho_{\rm I}$ and $\Delta \varphi_{\rm M} - \Delta \varphi_{\rm I}$ corresponding to these samples are plotted in Appendix \ref{sec:comp}. Only the beam intensities of both poles of this pulsar may be similar. Intrinsic emission region maps of this pulsar suggest that active emission regions at both poles are likely on the trailing side.

\subsection*{PSR J2047$+$5029}

This pulsar has sparse PAs due to the limited linear polarization intensity and temporal bins. Our A/R-corrected fit gives $\alpha = 161^\circ{}^{+8}_{-8}$ and $\beta = -0.8^\circ{}^{+0.3}_{-0.3}$, slightly different from \cite{Wang+2023raa}, possibly due to the large interpulse emission height. Despite the steep PA swing slopes, the fitted emission geometry suggests a single-pole configuration, and the fitted curve does not fully match the interpulse PA swing. Further observations are needed for a reliable PA swing.

\subsection*{PSR J2208$+$4056}

This pulsar shows strong linear polarization and a regular PA swing. \cite{Wang+2023raa} and \cite{Sun+2025apj} presented different treatments for the PA swing. \cite{Wang+2023raa} considered that the polarization mode of the interpulse is different from that of the main pulse and applied a $-90^\circ$ jump to the interpulse PA swing, while \cite{Sun+2025apj} applied no OPM jump. Following \cite{Sun+2025apj} caused our A/R-corrected RVM fit to fail to converge. Adopting the \cite{Wang+2023raa} treatment, we obtain $\alpha = 151^\circ{}^{+2}_{-2}$ and $\beta = -6.8^\circ{}^{+0.6}_{-0.6}$, differing from their results, and the A/R-corrected PA swing also does not fit the interpulse PAs well. This may be due to the large interpulse impact angle and the corresponding high emission altitude, possibly inconsistent with the polar cap model. A more comprehensive study on this pulsar is needed.

\section{Conclusions and Discussions}
\label{sec:conc&diss}

In this work, to investigate whether the emission beams from the two poles of pulsars are similar, we selected 11 pulsars that appear to have double-pole emission geometry observed with FAST, MeerKAT, and Parkes. Subsequently, we introduced an A/R-corrected RVM to fit their observed PA swings, aiming to determine the intrinsic emission regions at the two poles. Using intrinsic emission radii and azimuth angles derived from the intrinsic emission regions, we compared the emission beams from the two poles of eight pulsars with a confirmed double-pole emission geometry under both the conal beam and fan beam models. None shows similar magnetic azimuth widths. Only two pulsars have potentially similar maximum emission radii. In contrast, six pulsars show comparable beam intensities between their two poles. 

The observed differences in beam sizes imply that physical conditions in the pair creation regions are not identical at the two magnetic poles of a pulsar. Moreover, as summarized in Table \ref{tab:comp}, the active emission regions of these pulsars may be distributed randomly on the two poles. The random distribution and unfilled nature further indicate inhomogeneity in the active magnetic flux tubes within the polar cap region. These asymmetries could originate from differences in the local magnetic field structure \citep[e.g.,][]{Gil+2006ApJ, Szary2013arXiv1304.4203S} or surface properties at each pole \citep[e.g.,][]{Xu+2026RAA}.

A key assumption of our A/R-corrected RVM is that radio emission originates at low altitudes in a static dipole. Although RVM for a multipole field can be identical to the dipole case \citep{Qiu+2023ApJ}, emission heights would be lower, reducing A/R phase shifts. For pulsars with very low emission heights ($< 0.1 R_{\rm LC}$), this scenario may hold. However, the emission heights are substantial in some cases (pulsars with large impact angles or single-pole emission geometry), suggesting that radio emission could originate at high altitudes, as previously proposed \citep{Romani+1995ApJ, Muslimov+2003ApJ}. Moreover, the A/R-corrected RVM curve can be distorted at high altitudes (e.g., PSR J1755$-$0903), requiring more comprehensive modeling.

Another key assumption is that the observed radio emission originates from field lines with a specific footprint parameter. Moreover, the uncertainties in $\eta$ for each pole of selected pulsars are relatively small. This does not imply that the observed radio emission arises exclusively from a single layer of the field lines. In this study, we considered only mean pulse profiles, and the fitted $\eta$ values are representative of these averaged profiles. The PAs from single pulses may scatter around the mean PAs \citep[e.g.,][]{Gil+1995MNRAS, Rankin+1995JApA, Ramachandran+2004ApJ, Mitra+2016MNRAS, Johnston+2024MNRAS}, suggesting that radio emission could also arise from other layers of the field lines.


\begin{acknowledgments}
We are grateful to B. Posselt for sharing the code and offering guidance on generating noiseless pulse profiles with the Gaussian process method. We also thank A. Karastergiou for making the open-source \textsc{Pulseshape} algorithm for Gaussian processes publicly available. This work is supported by the National Natural Science Foundation of China grant 12133004 and the National SKA Program of China (Nos. 2020SKA0120101, 2020SKA0120300). This work is supported by Guangdong S\&T programme No. 2024B1212080003. We would like to thank all the anonymous reviewers for their valuable comments and assistance.
\end{acknowledgments}

\bibliography{sample631}{}
\bibliographystyle{aasjournal}

%

\vspace{5mm}




\clearpage
\appendix
\section{The inner open field lines and coordinate transformation}
\label{sec:field&trans}

\restartappendixnumbering

This appendix derives the analytic equation for inner field lines whose footprints are uniformly distributed on the stellar surface and provides the transformation matrices between coordinate systems.

\begin{figure}
    \centering
    \includegraphics[width=0.5\linewidth]{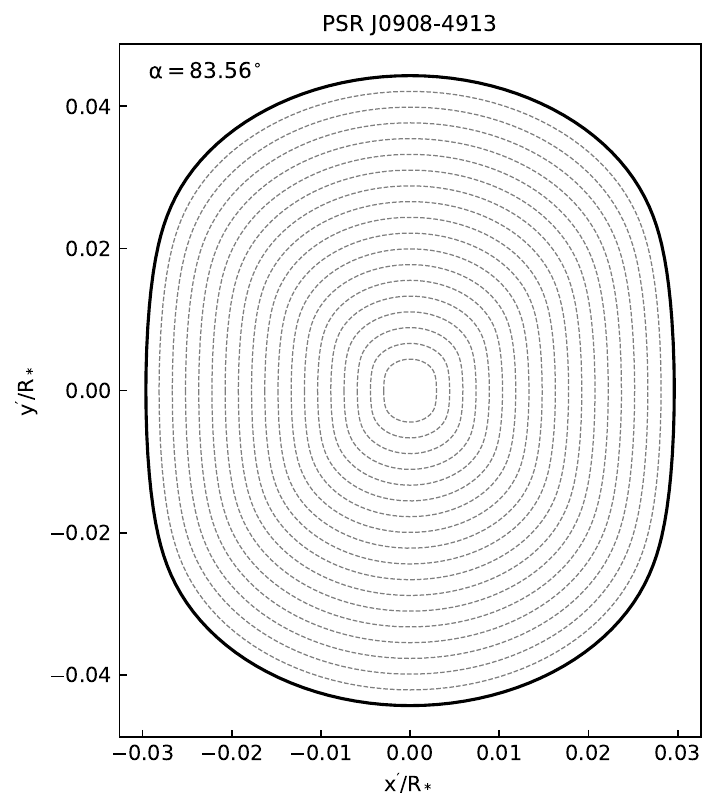}
    \caption{Example footprints on the stellar surface for the last open field lines (black solid) and inner open field lines (gray dashed).}
    \label{fig:fp}
\end{figure}

On the stellar surface, the last open field lines described by $h_{\rm em} = R_{\rm e} \sin^2 \theta$, satisfy 
\begin{equation}
    \label{equ:lastfp}
    R_*=R_{\rm e}\sin^2 \theta_{\rm pc} \,,
\end{equation}
where $R_*$ and $\theta_{\rm pc}$ are the stellar radius and the polar angle of the edge of the polar cap, respectively. Changing the polar angle to $\eta \theta_{\rm pc}$ requires a coefficient $\iota$ to keep the altitude equal to $R_*$. The equation for inner footprints becomes
\begin{equation}
    \label{equ:innerfp}
    R_*=\iota R_{\rm e}\sin^2 (\eta \theta_{\rm pc}) \,.
\end{equation}
Equation \ref{equ:inneropen} is then obtained from $h_{\rm em}=\iota R_{\rm e}\sin^2\theta$, with $\iota$ derived from Equation \ref{equ:lastfp} and \ref{equ:innerfp}. The footprints of the last open field lines and inner field lines are shown in Figure \ref{fig:fp}.

Regarding coordinate transformations, vectors $\mathbf{k}'_{\rm m}$ and $\mathbf{r}'_{\rm m}$ derived from emission longitudes are initially in spherical coordinates (see Section \ref{subsec:ar}). They are converted to Cartesian coordinates via
\begin{equation}
    \left(
    \begin{matrix}
        x \\
        y \\
        z 
    \end{matrix}        
    \right)
    =
    r
    \left(
    \begin{matrix}
        \sin\Theta\cos\Phi\\
        \sin\Theta\sin\Phi\\
        \cos\Theta
    \end{matrix}
    \right),
\end{equation}
where $x$, $y$, and $z$ are the components along the unit vectors of $\mathcal{K}'_{\rm m}$, and $r$, $\Theta$, and $\Phi$ are the spherical coordinates of the vector.

Following \cite{2010MNRAS.405.2103L}, the transformation from $\mathcal{K}'_{\rm m}$ to $\mathcal{K}'$ is $\mathbf{a}=\mathbf{T}\mathbf{a}_{\rm m}$, with
\begin{equation}
    \mathbf{T}=\left(\begin{matrix}{}
        \cos\alpha & 0 & \sin\alpha \\
        0 & 1 & 0 \\
        -\sin\alpha & 0 & \cos\alpha \\
    \end{matrix}
    \right),
\end{equation} 
where $\mathbf{a}$ and $\mathbf{a}_{\rm m}$ denote the vectors $\mathbf{k}$ or $\mathbf{r}$ in $\mathcal{K}'$ and $\mathcal{K}'_{\rm m}$, respectively.

The transformation of $\mathbf{k}$ from $\mathcal{K}'$ to $\mathcal{K}''$ used in Section \ref{subsec:ar} is $\mathbf{k}''=\mathbf{C}\mathbf{k}'$, with
\begin{equation}
    \mathbf{C}=\left(
    \begin{matrix}
        \cos \phi' & \sin \phi' & 0 \\
        -\sin \phi' & \cos \phi' & 0 \\
        0 & 0 & 1
    \end{matrix}
    \right),
\end{equation}
where $\phi' = \phi_{\rm r} + 90^\circ$. Conversely, $\mathbf{k}_{\rm a}''$ is transformed to $\mathbf{k}'_{\rm a}$ via $\mathbf{k}'_{\rm a}=\mathbf{C}^{-1}\mathbf{k}''_{\rm a}$, where $\mathbf{C}^{-1}=\mathbf{C}^{\rm T}$ because $\mathbf{C}$ is a rotation matrix.

\clearpage

\section{Fitted Results with the A/R-corrected RVM}
\label{sec:prof}

\restartappendixnumbering

\begin{figure*}[h]
\centering
    \subfigure{
    \centering
    \includegraphics[width=0.85\linewidth]{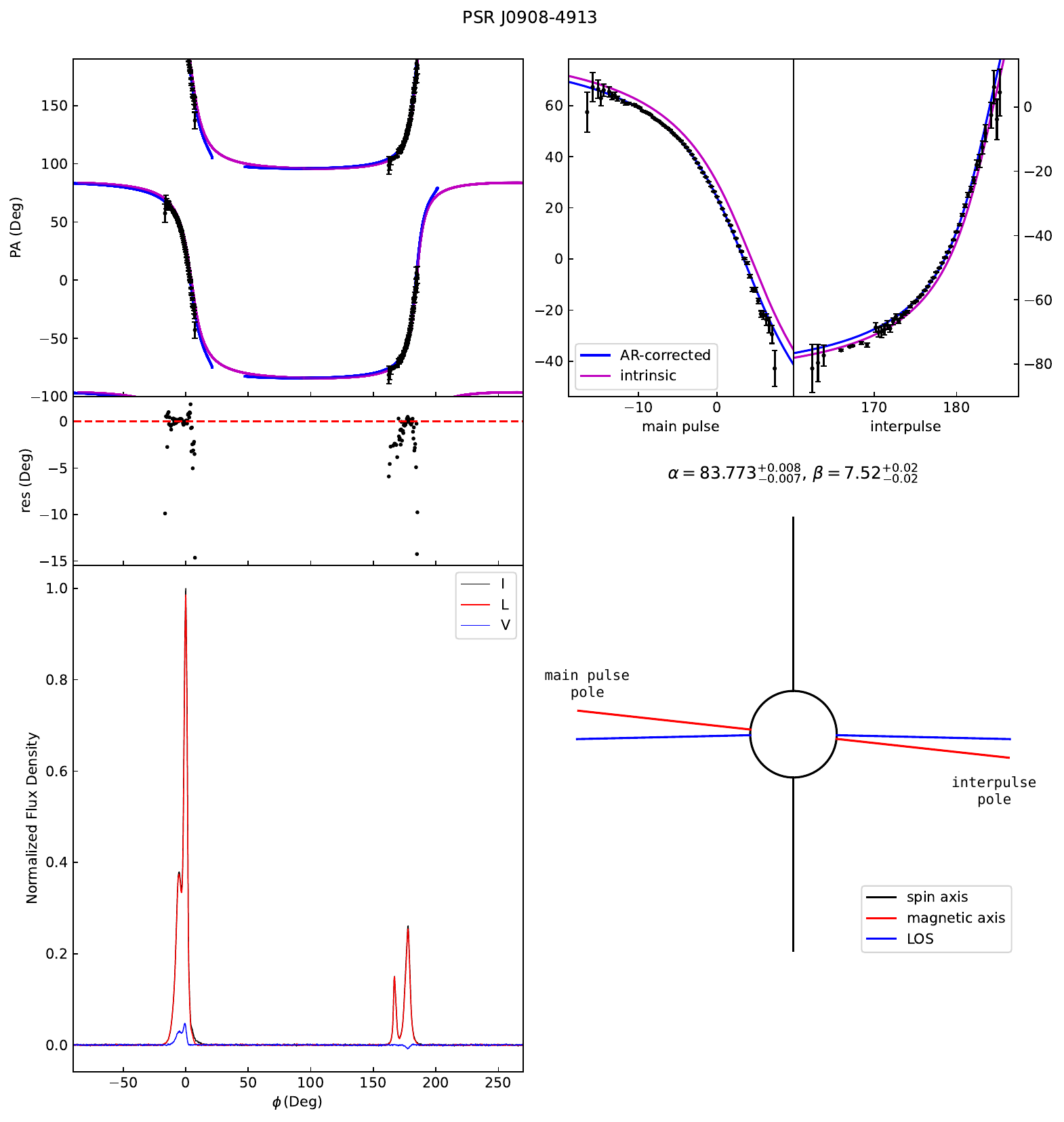}
    }
    \caption{An example of the fitted result and emission geometry for PSR J0908$-$4913. Left panels, from top to bottom: observed PAs (black, blue, and gray with error bars), the A/R-corrected PA swings (blue), and the intrinsic PA swings (magenta), the fitting residuals, and the mean polarization profile (black for total intensity, red for linear polarization, blue for circular polarization). Gray PAs are excluded from the fit. Blue PAs are data with manually applied $90^\circ$ jumps. A/R-corrected and intrinsic PA swings from 200 random samples within $1\sigma$ credible interval are plotted to visualize the fitting uncertainty. The left half of the upper right panel zooms in on observed PAs and fitted curves corresponding to the median values for the main pulse; the right half shows the same for the interpulse. The lower right panel shows the schematic emission geometry. The left half shows the relative position between the main pulse pole and the LOS as it sweeps across the main pulse pole; the right half shows the relative position between the interpulse pole and the LOS as it sweeps across the interpulse pole.}
    \label{fig:prof0908}
\end{figure*}

\begin{figure*}[h]
\centering
    \subfigure{
    \centering
    \includegraphics[width=\linewidth]{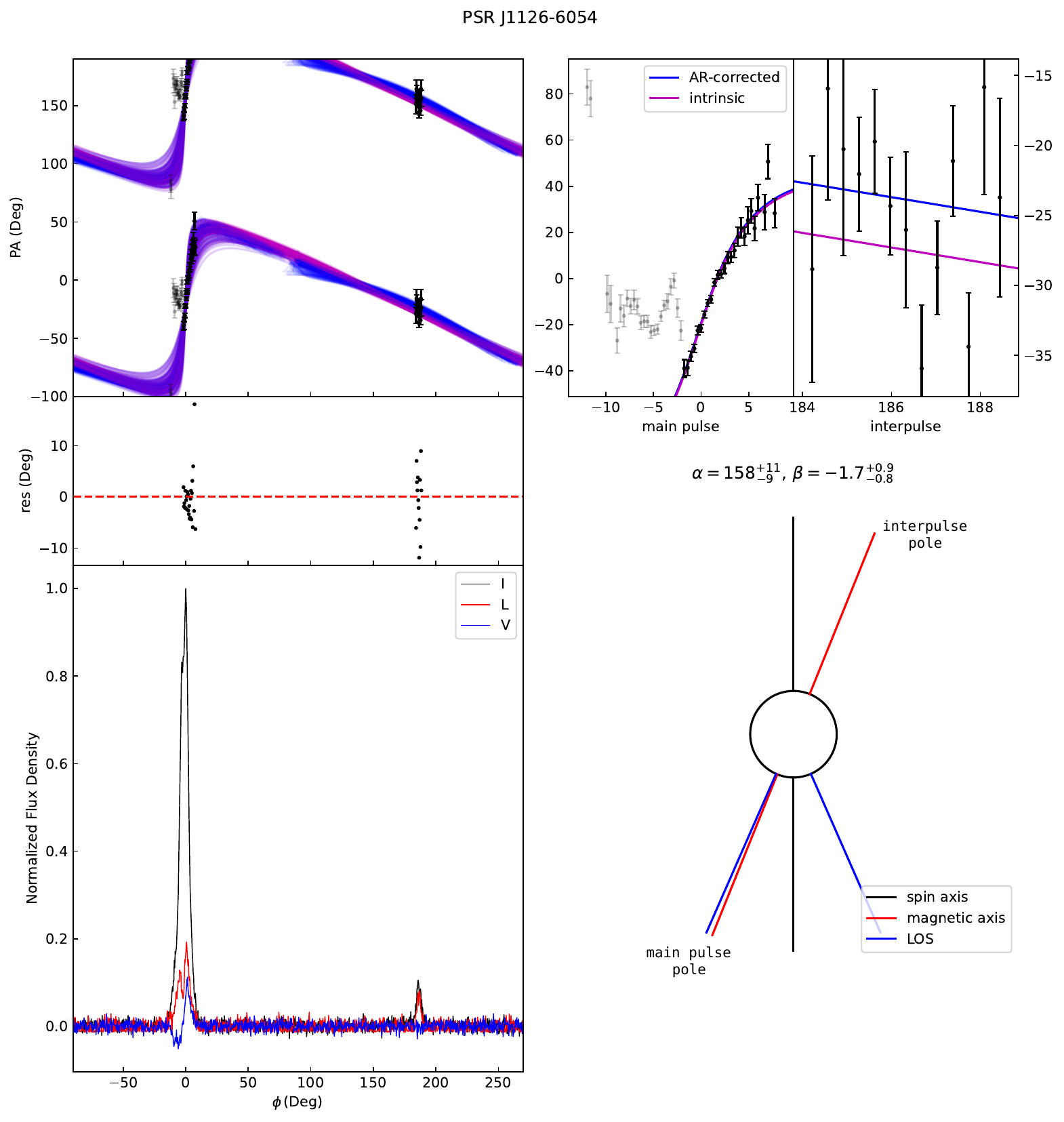}
    }
    \caption{The same as Figure \ref{fig:prof0908} but for PSR J1126$-$6054.}
    \label{fig:prof1126}
\end{figure*}

\begin{figure*}[h]
\centering
    \subfigure{
    \centering
    \includegraphics[width=\linewidth]{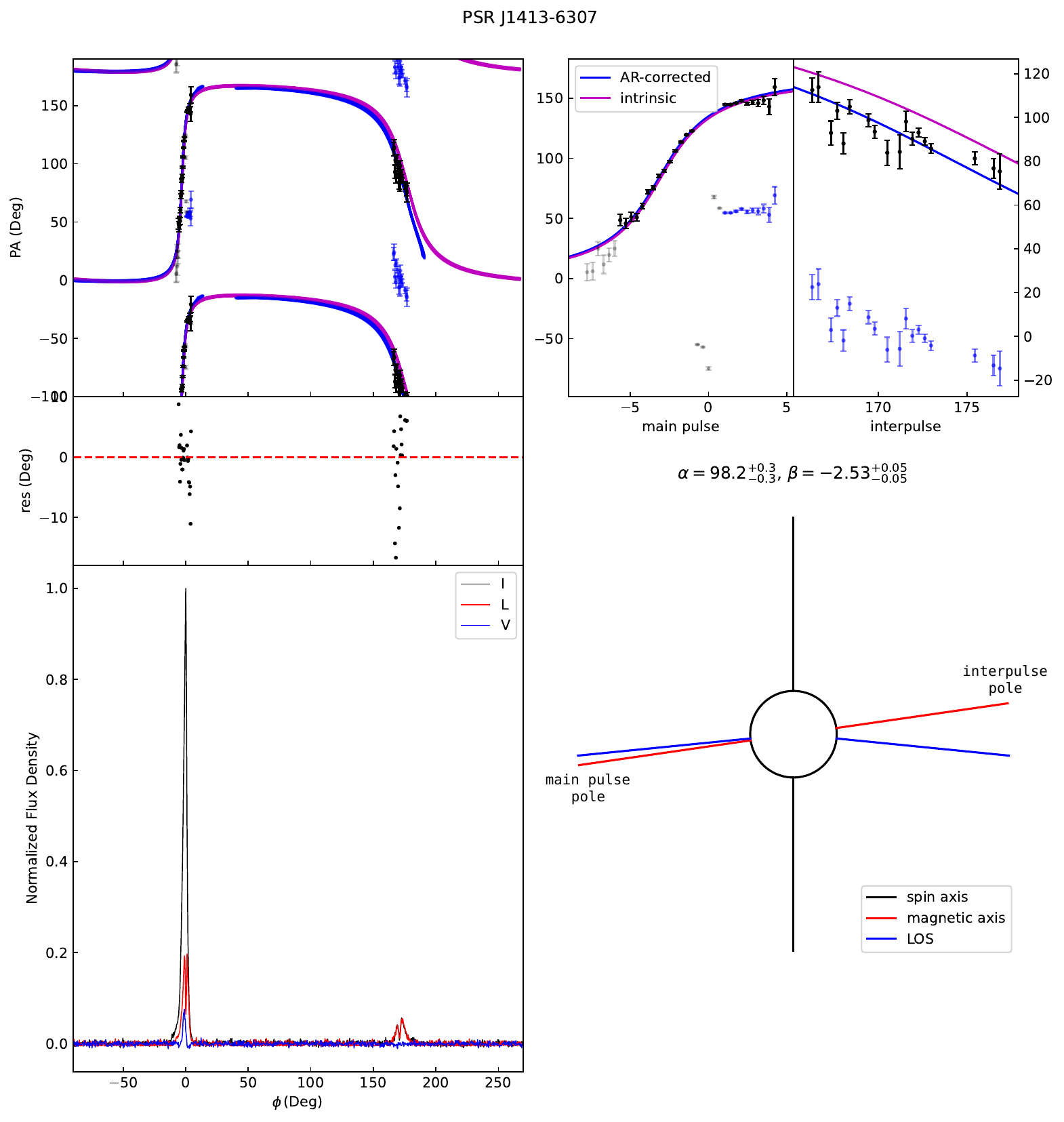}
    }
    \caption{The same as Figure \ref{fig:prof0908} but for PSR J1413$-$6307.}
    \label{fig:prof1413}
\end{figure*}

\begin{figure*}[h]
\centering
    \subfigure{
    \centering
    \includegraphics[width=\linewidth]{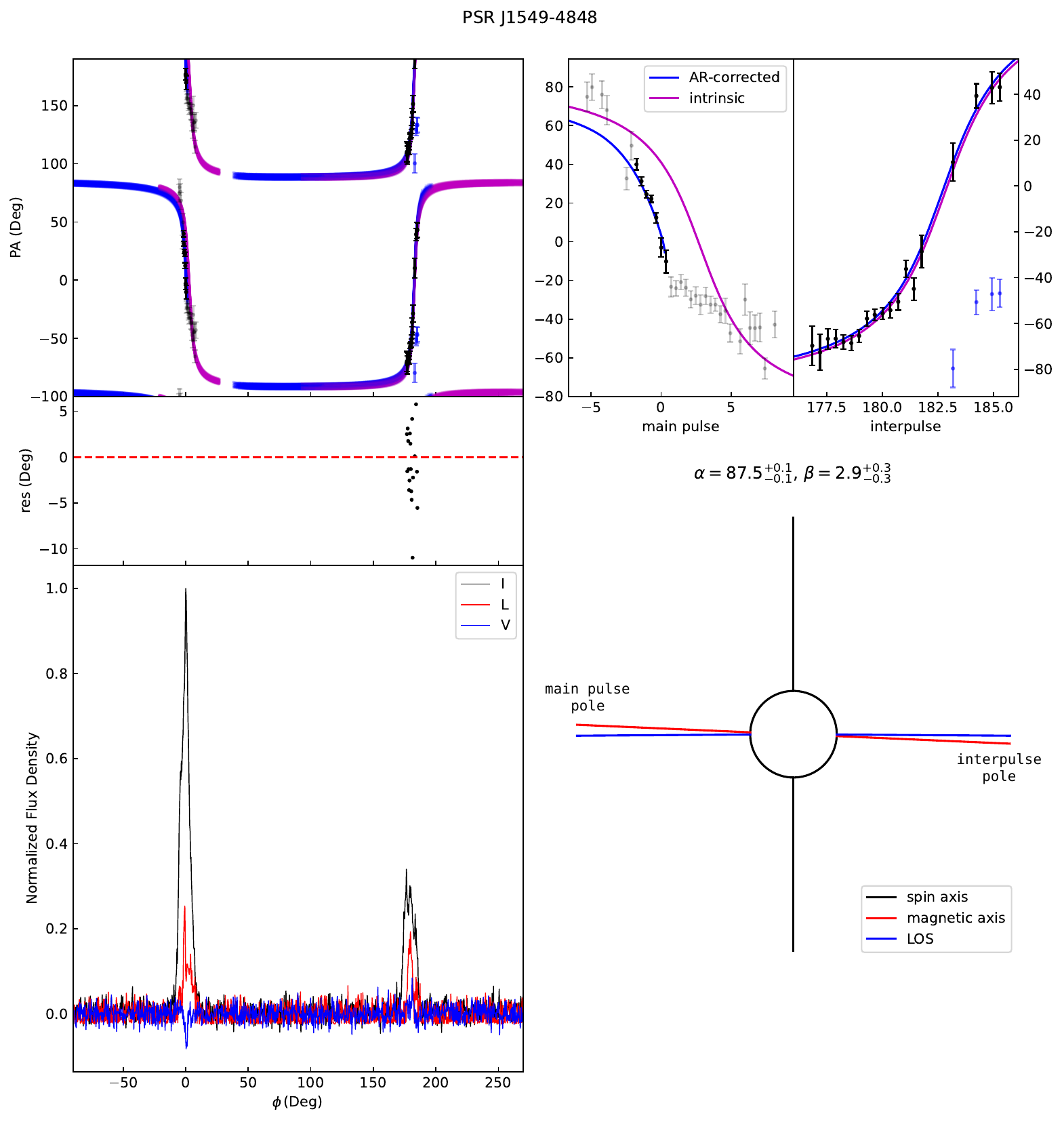}
    }
    \caption{The same as Figure \ref{fig:prof0908} but for PSR J1549$-$4848.}
    \label{fig:prof1549}
\end{figure*}

\begin{figure*}[h]
\centering
    \subfigure{
    \centering
    \includegraphics[width=\linewidth]{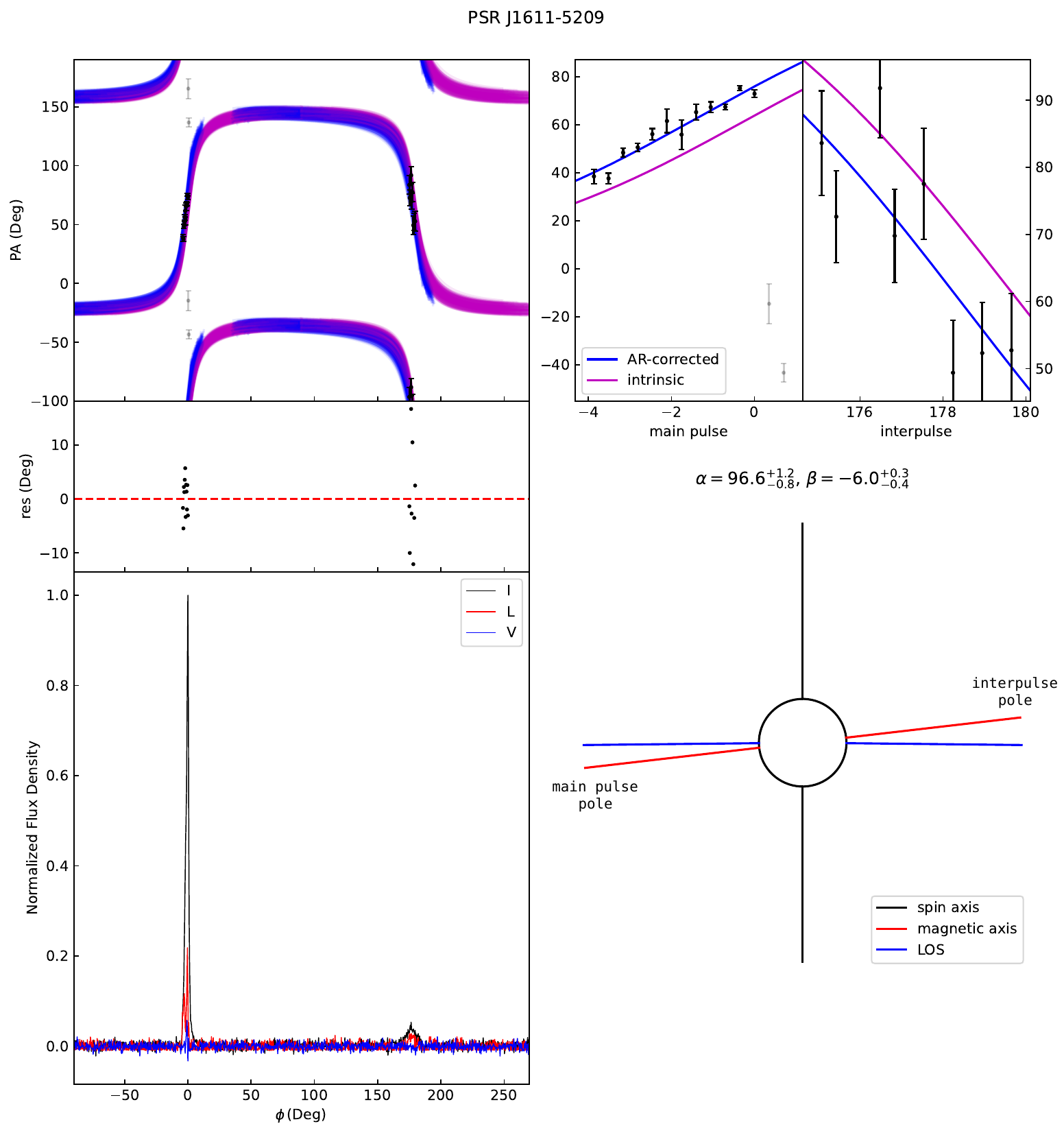}
    }
    \caption{The same as Figure \ref{fig:prof0908} but for PSR J1611$-$5209.}
    \label{fig:prof1611}
\end{figure*}

\begin{figure*}[h]
\centering
    \subfigure{
    \centering
    \includegraphics[width=\linewidth]{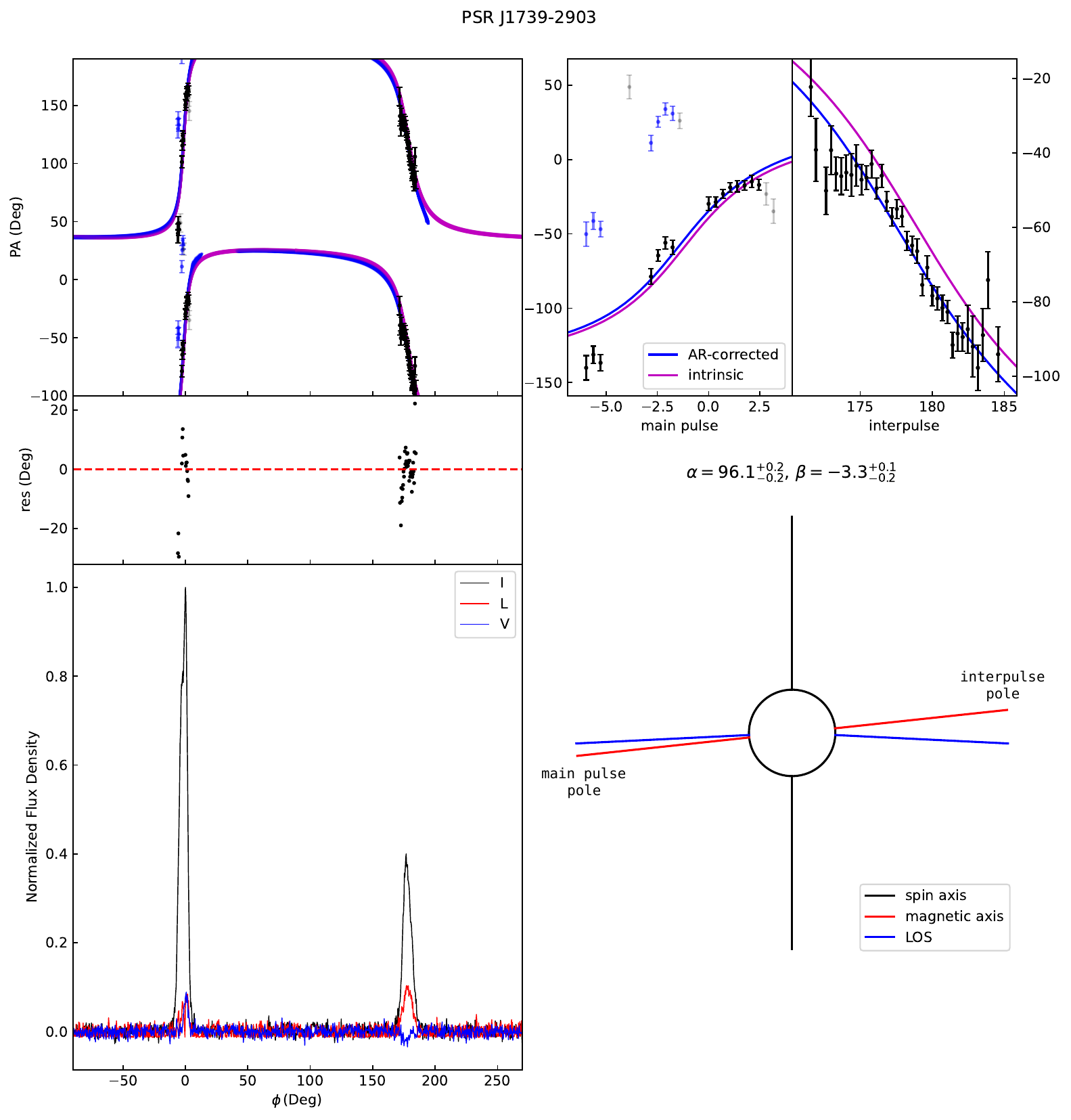}
    }
    \caption{The same as Figure \ref{fig:prof0908} but for PSR J1739$-$2903.}
    \label{fig:prof1755}
\end{figure*}

\begin{figure*}[h]
\centering
    \subfigure{
    \centering
    \includegraphics[width=\linewidth]{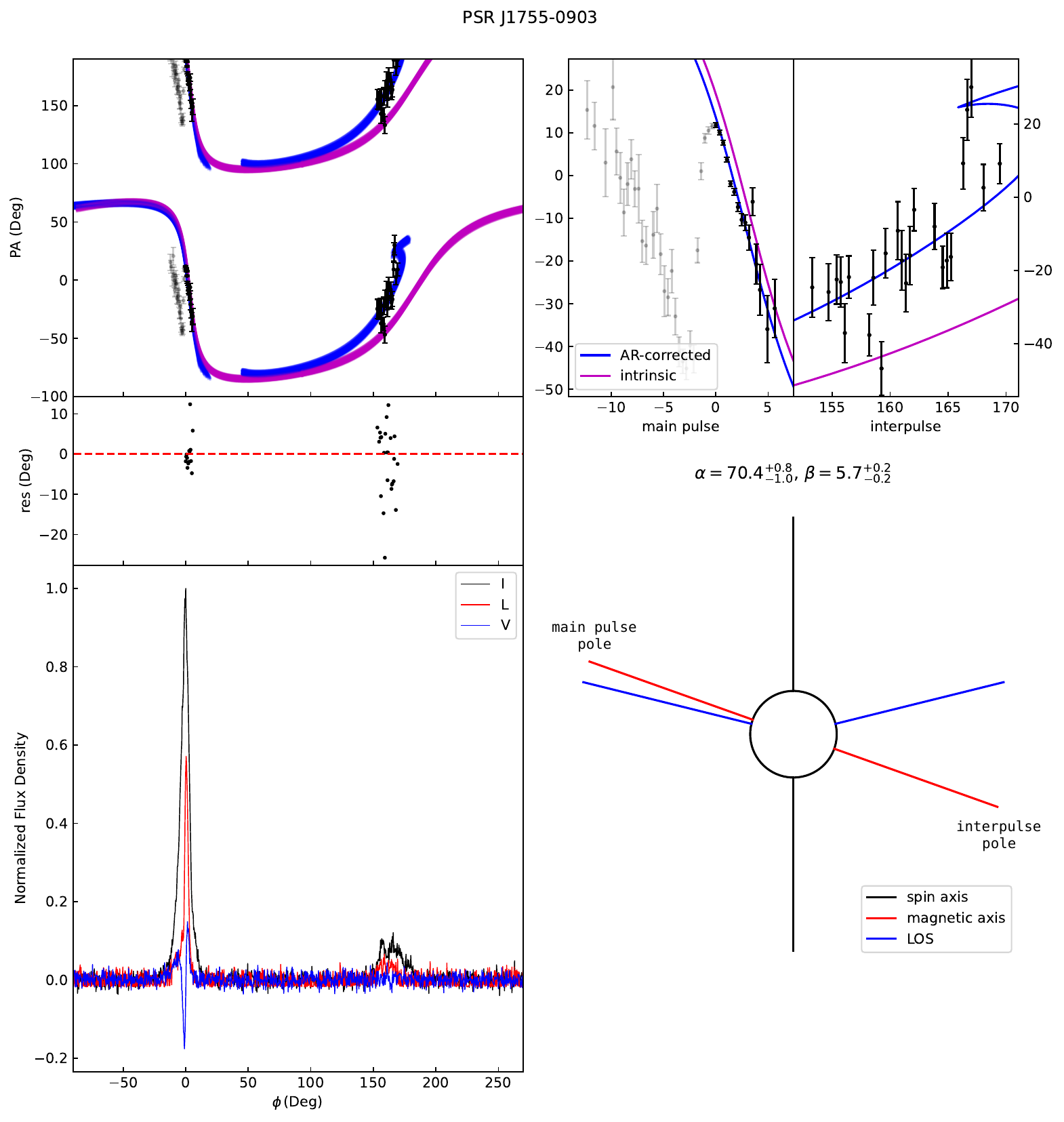}
    }
    \caption{The same as Figure \ref{fig:prof0908} but for PSR J1755$-$0903.}
    \label{fig:prof1755}
\end{figure*}

\begin{figure*}[h]
\centering
    \subfigure{
    \centering
    \includegraphics[width=\linewidth]{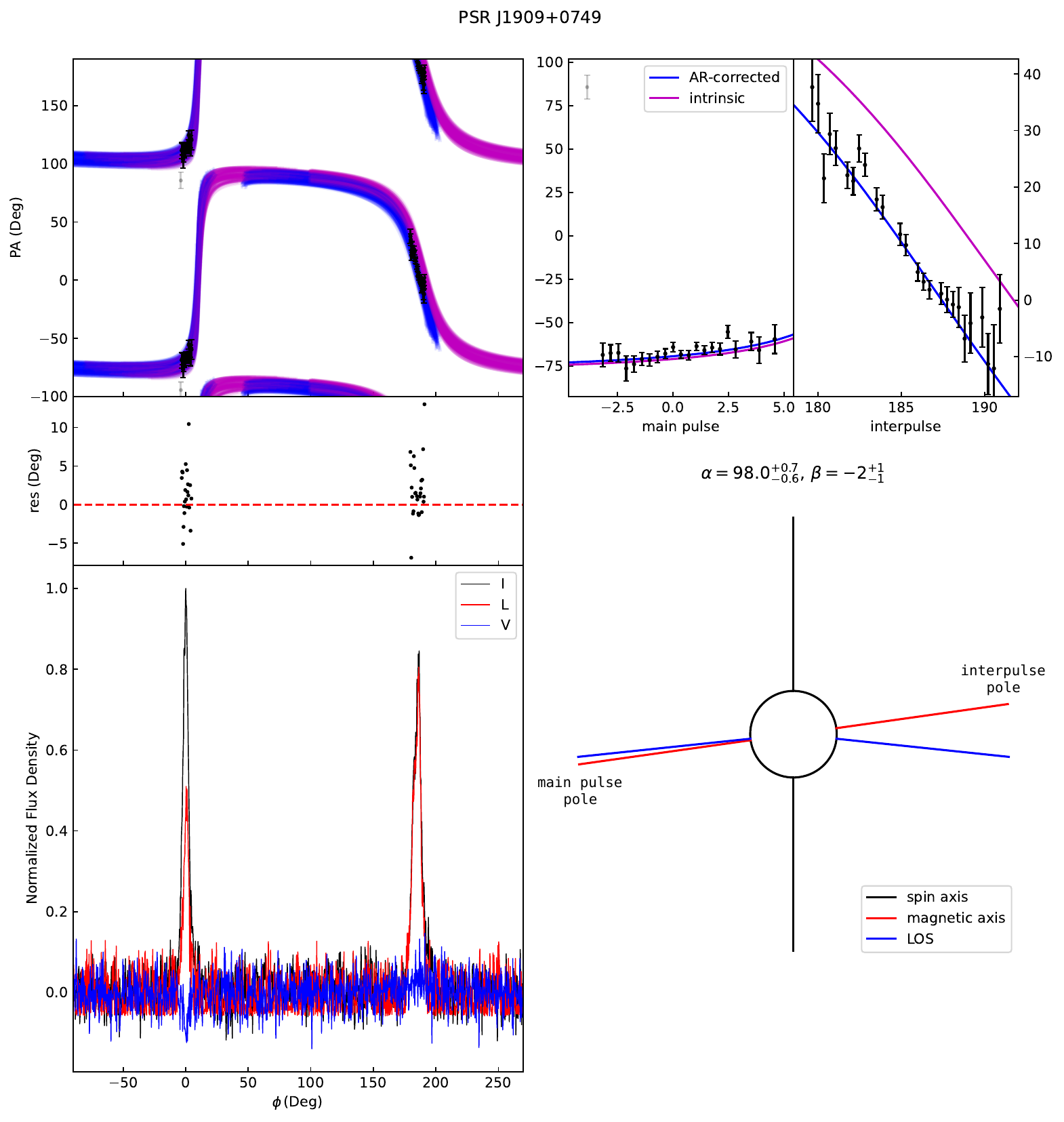}
    }
    \caption{The same as Figure \ref{fig:prof0908} but for PSR J1909$+$0749.}
    \label{fig:prof1909}
\end{figure*}

\begin{figure*}[h]
\centering
    \subfigure{
    \centering
    \includegraphics[width=\linewidth]{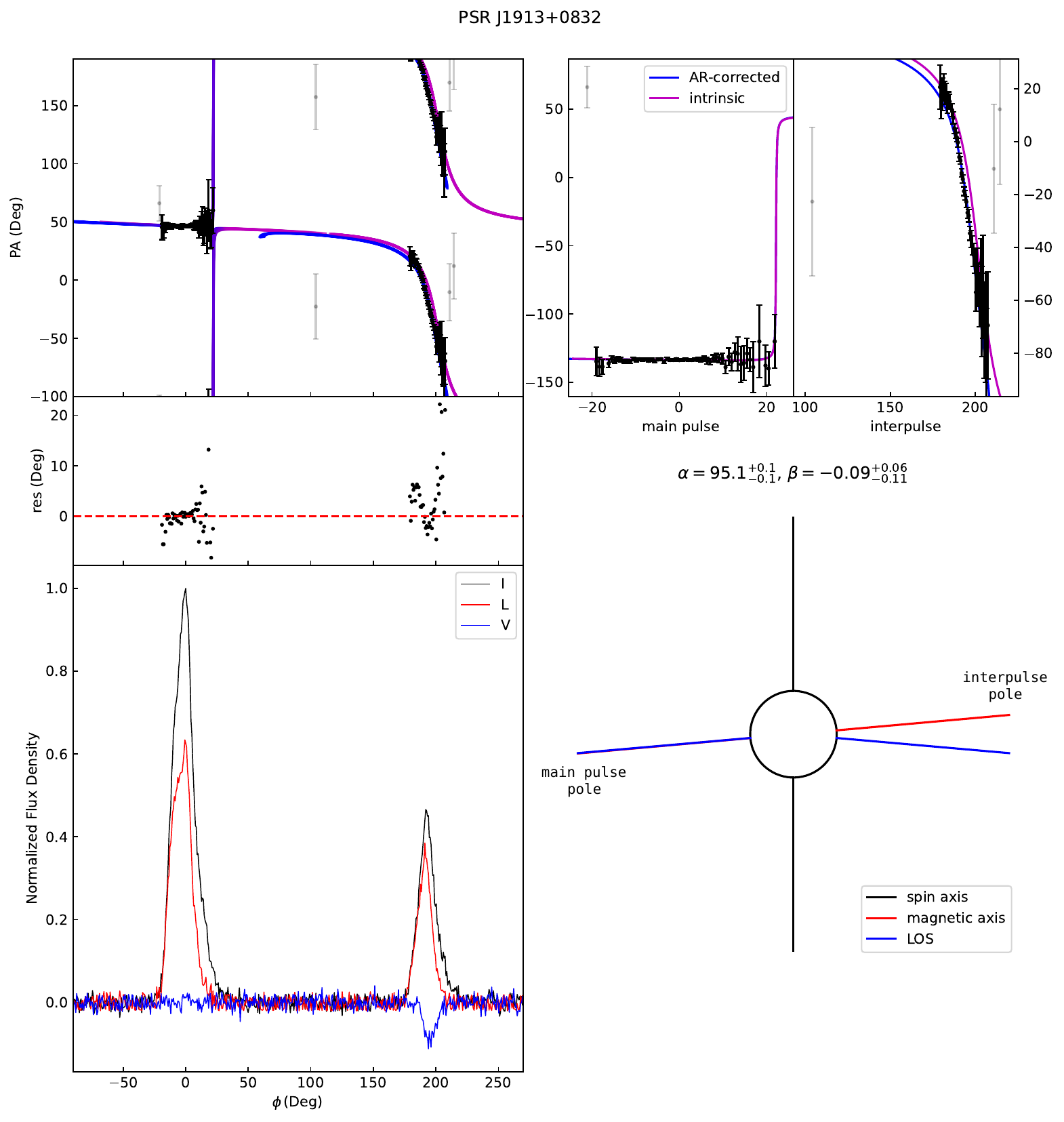}
    }
    \caption{The same as Figure \ref{fig:prof0908} but for PSR J1913$+$0832.}
    \label{fig:prof1913}
\end{figure*}

\begin{figure*}[h]
\centering
    \subfigure{
    \centering
    \includegraphics[width=\linewidth]{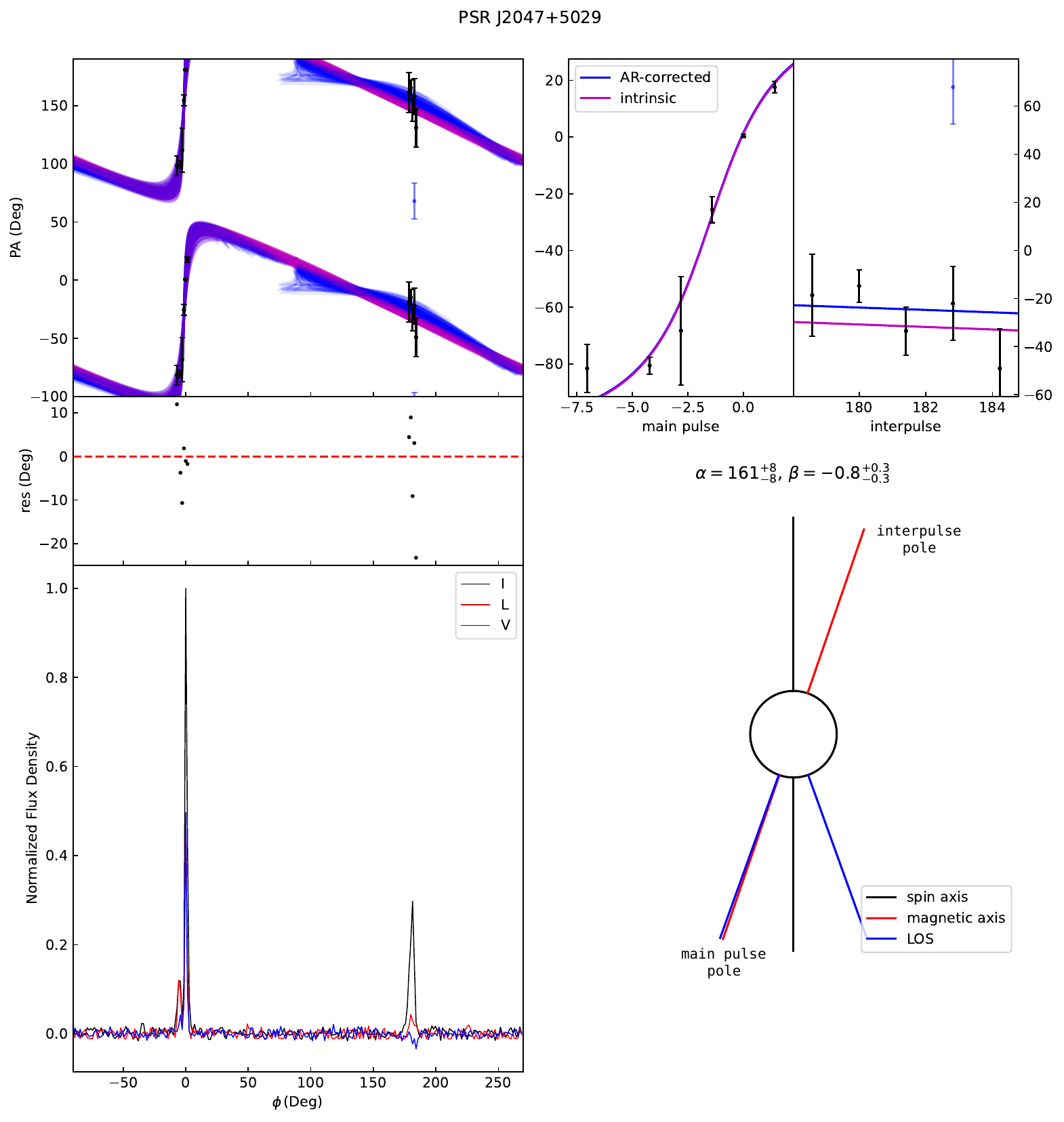}
    }
    \caption{The same as Figure \ref{fig:prof0908} but for PSR J2047$+$5029.}
    \label{fig:prof2047}
\end{figure*}

\begin{figure*}[h]
\centering
    \subfigure{
    \centering
    \includegraphics[width=\linewidth]{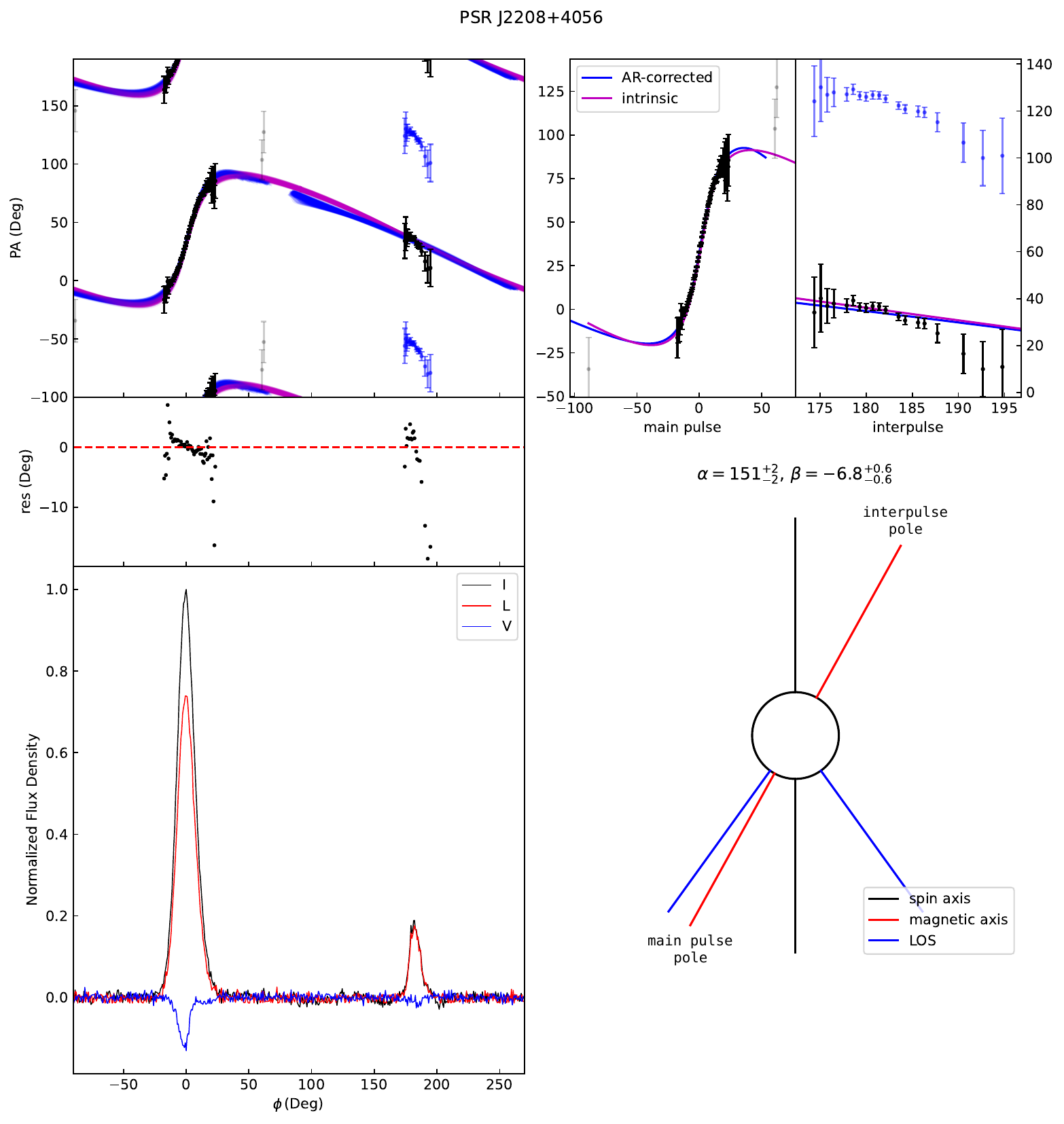}
    }
    \caption{The same as Figure \ref{fig:prof0908} but for PSR J2208$+$4056.}
    \label{fig:prof2208}
\end{figure*}

\clearpage

\section{Comparison Results}
\label{sec:comp}

\restartappendixnumbering

\begin{figure*}[h]
    \centering
    \subfigure{
    \centering
    \includegraphics[width=0.65\textwidth]{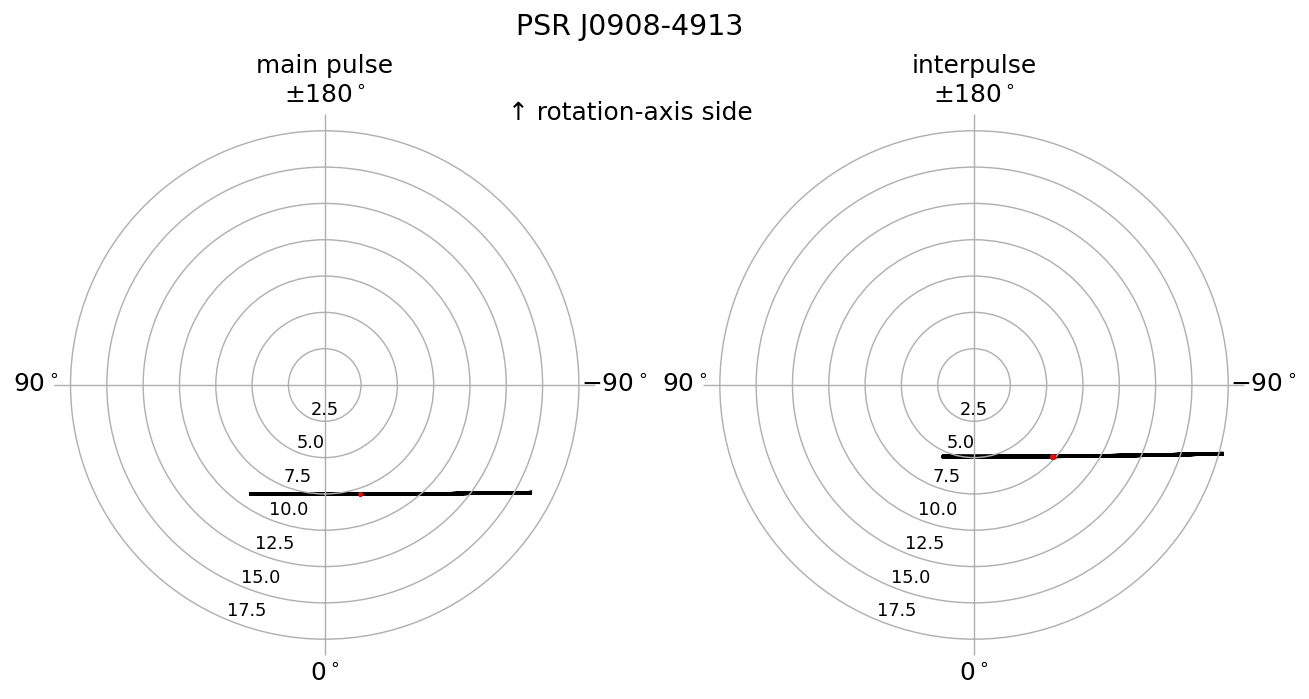}
    }
    \subfigure{
    \centering
    \includegraphics[width=0.65\textwidth]{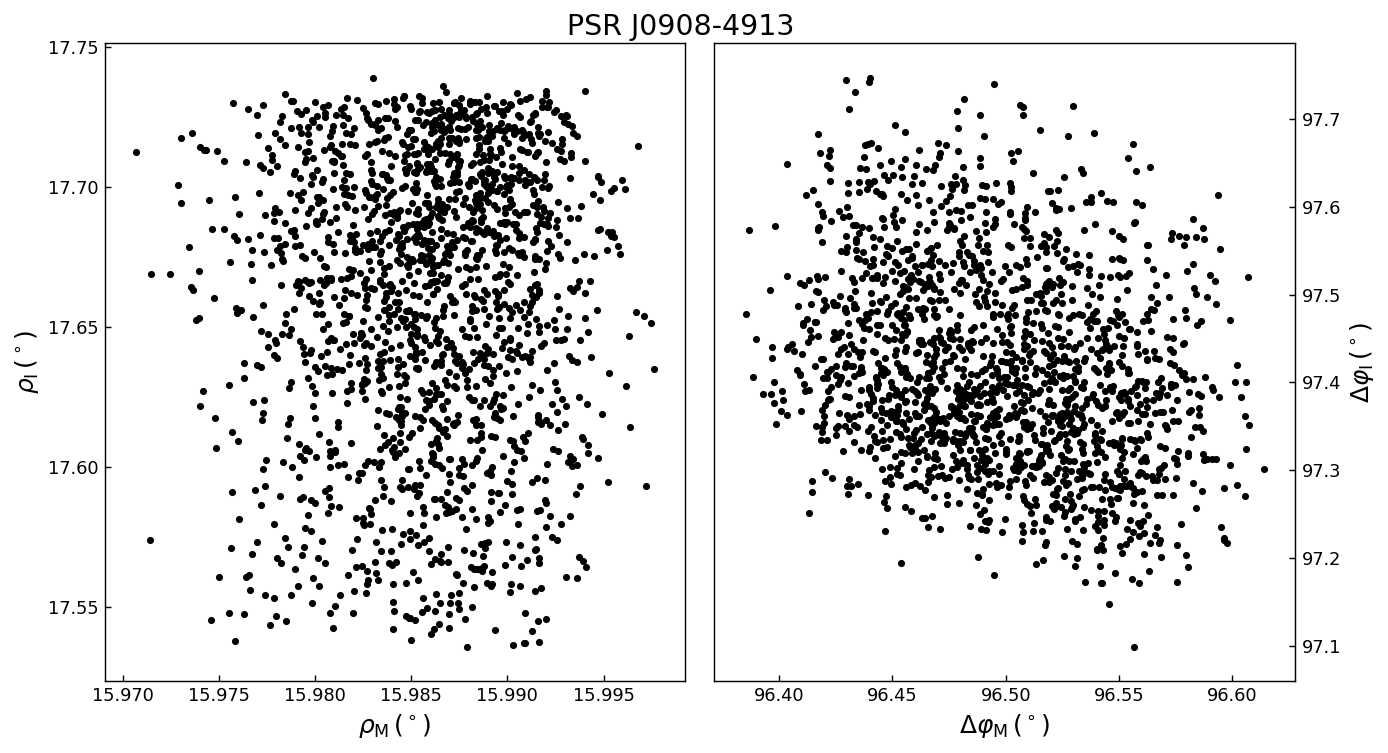}
    }
    \subfigure{
    \centering
    \includegraphics[width=0.35\textwidth]{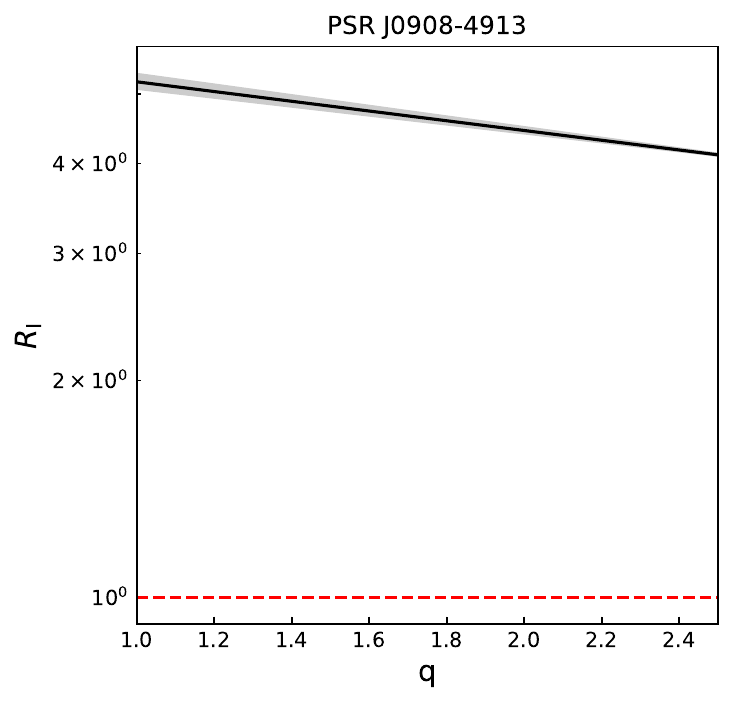}
    }
    \caption{Examples of the mapping plots, joint distributions, and $R_{\rm I}-q$ plots for PSR J0908$-$4913. See the keys in Figure \ref{fig:map}, \ref{fig:dist}, and \ref{fig:R_I-q}.}
    \label{fig:comp0908}
\end{figure*}

\begin{figure*}[h]
    \centering
    \subfigure{
    \centering
    \includegraphics[width=0.65\textwidth]{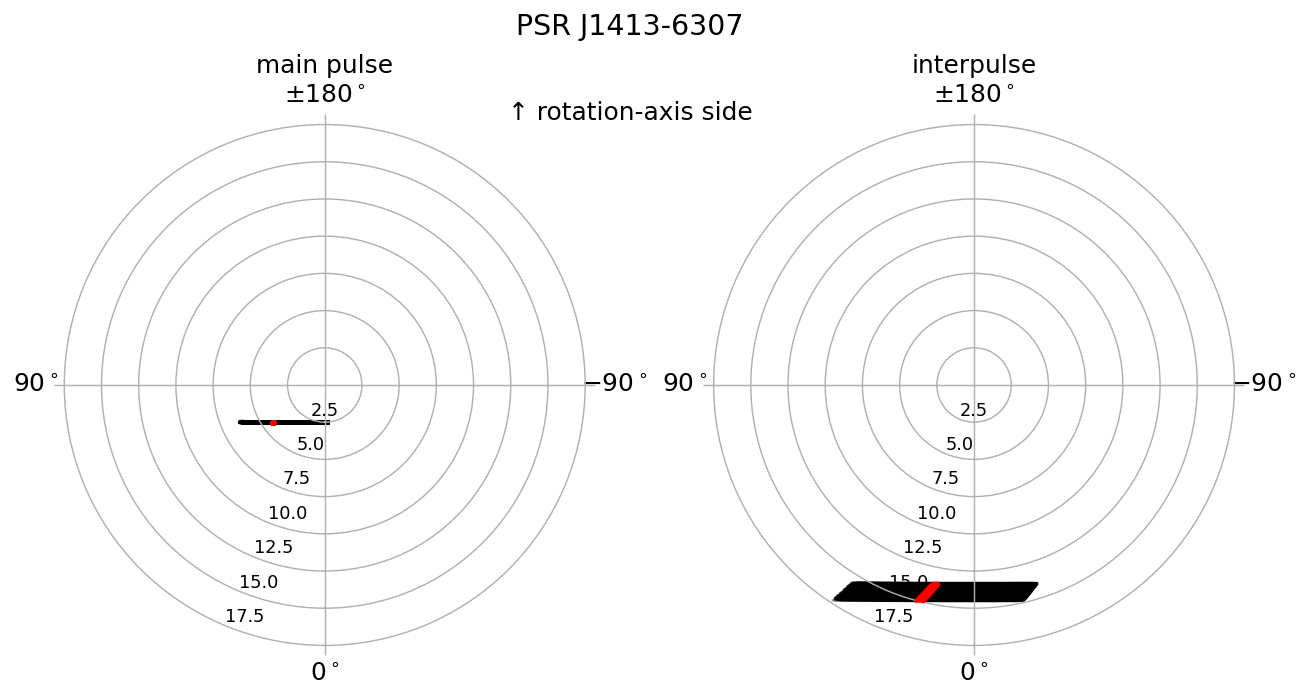}
    }
    \subfigure{
    \centering
    \includegraphics[width=0.65\textwidth]{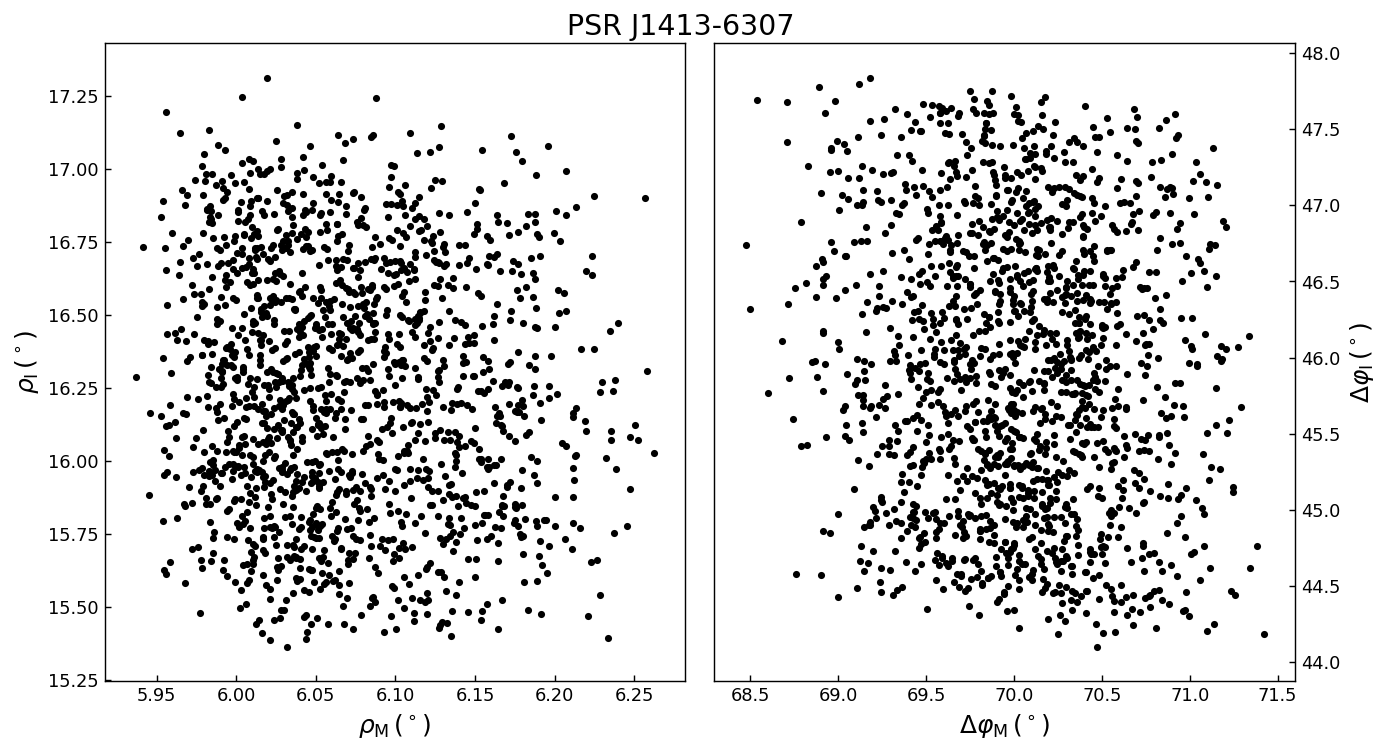}
    }
    \subfigure{
    \centering
    \includegraphics[width=0.35\textwidth]{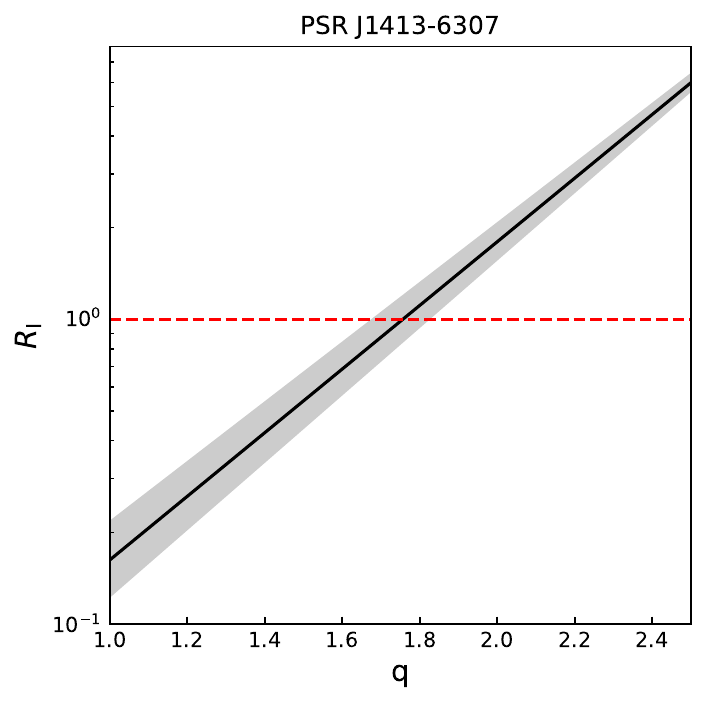}
    }
    \caption{The same as Figure \ref{fig:comp0908}, but for PSR J1413$-$6307.}
    \label{fig:comp1413}
\end{figure*}

\begin{figure*}[h]
    \centering
    \subfigure{
    \centering
    \includegraphics[width=0.65\textwidth]{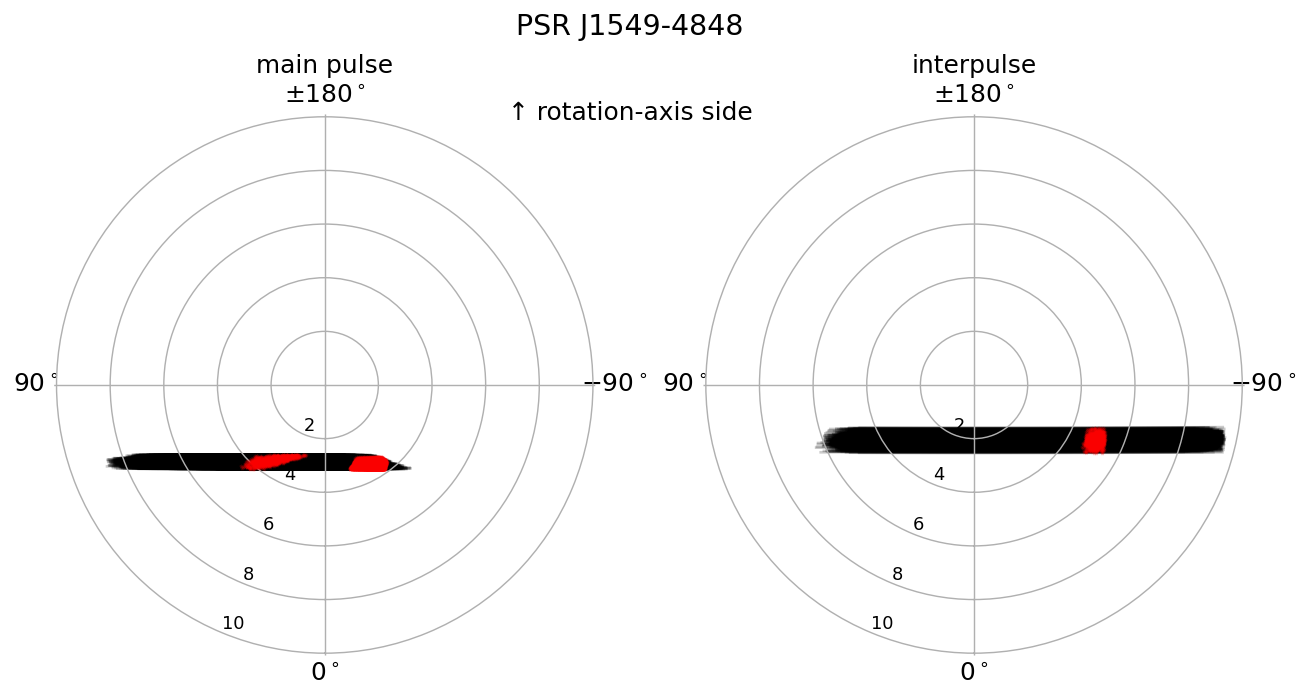}
    }
    \subfigure{
    \centering
    \includegraphics[width=0.65\textwidth]{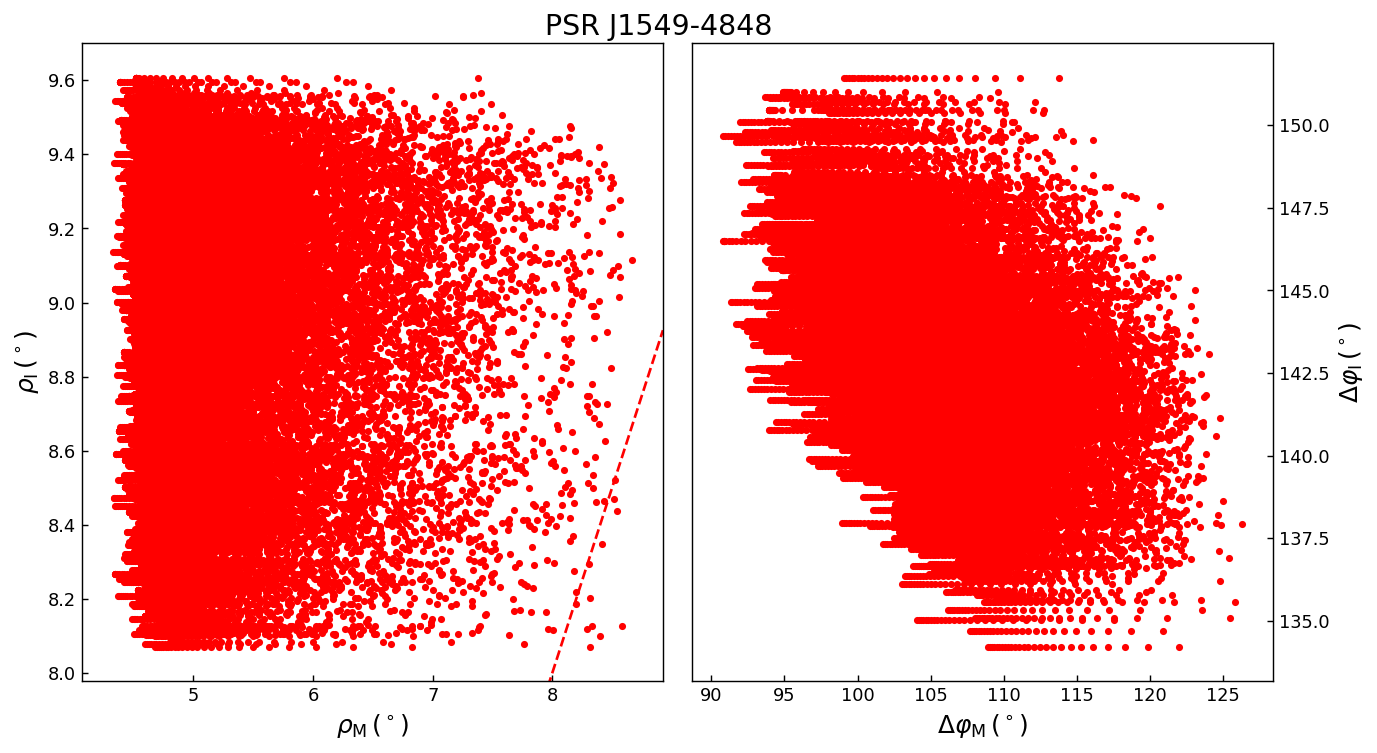}
    }
    \subfigure{
    \centering
    \includegraphics[width=0.35\textwidth]{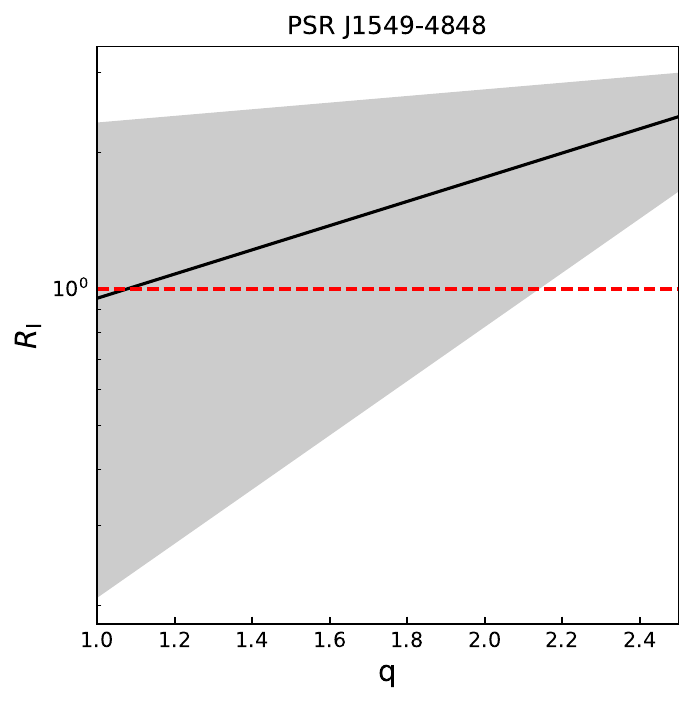}
    }
    \caption{The same as Figure \ref{fig:comp0908}, but for PSR J1549$-$4848.}
    \label{fig:comp1549}
\end{figure*}

\begin{figure*}[h]
    \centering
    \subfigure{
    \centering
    \includegraphics[width=0.65\textwidth]{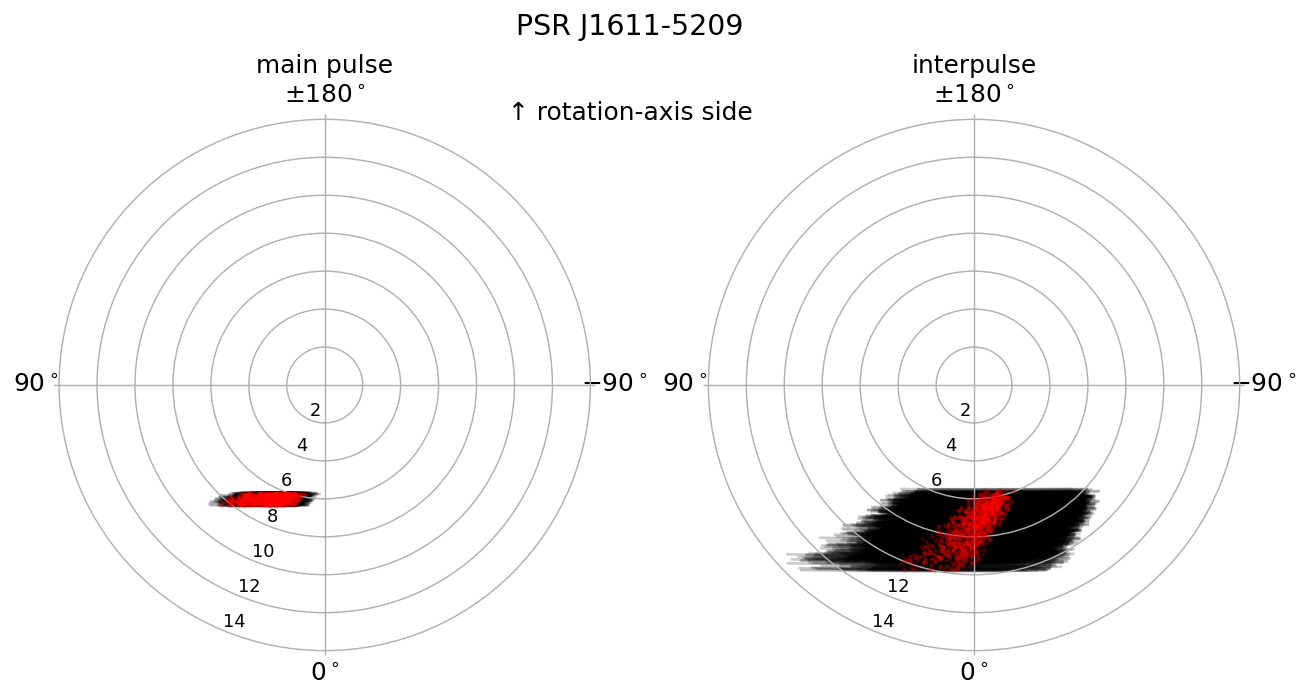}
    }
    \subfigure{
    \centering
    \includegraphics[width=0.65\textwidth]{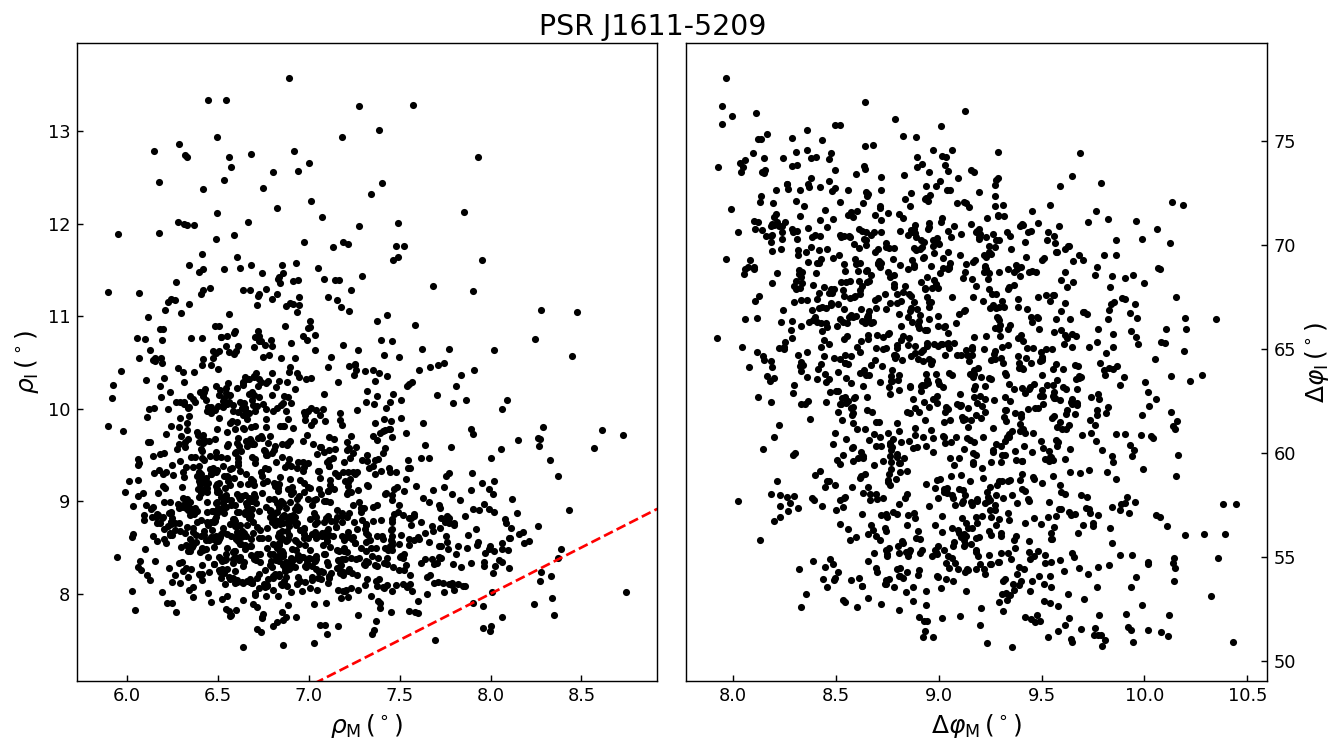}
    }
    \subfigure{
    \centering
    \includegraphics[width=0.35\textwidth]{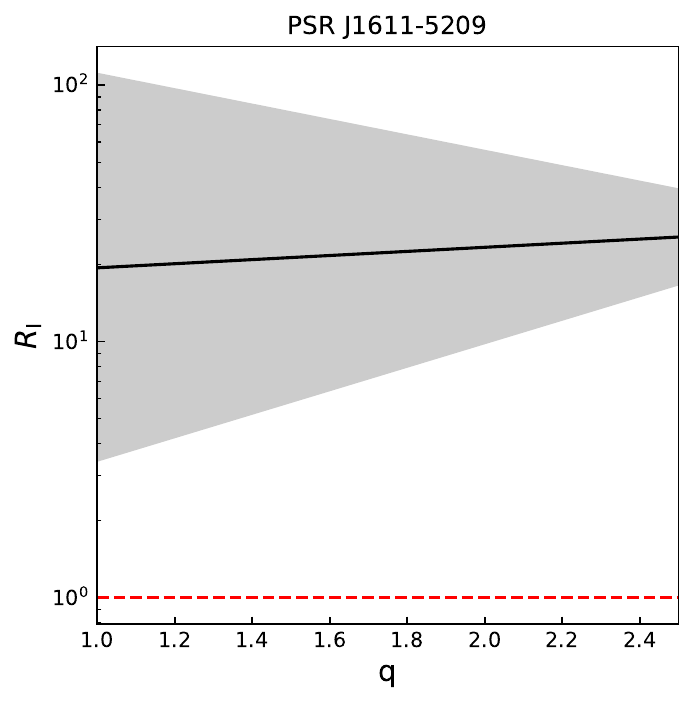}
    }
    \caption{The same as Figure \ref{fig:comp0908}, but for PSR J1611$-$5209.}
    \label{fig:comp1611}
\end{figure*}

\begin{figure*}[h]
    \centering
    \subfigure{
    \centering
    \includegraphics[width=0.65\textwidth]{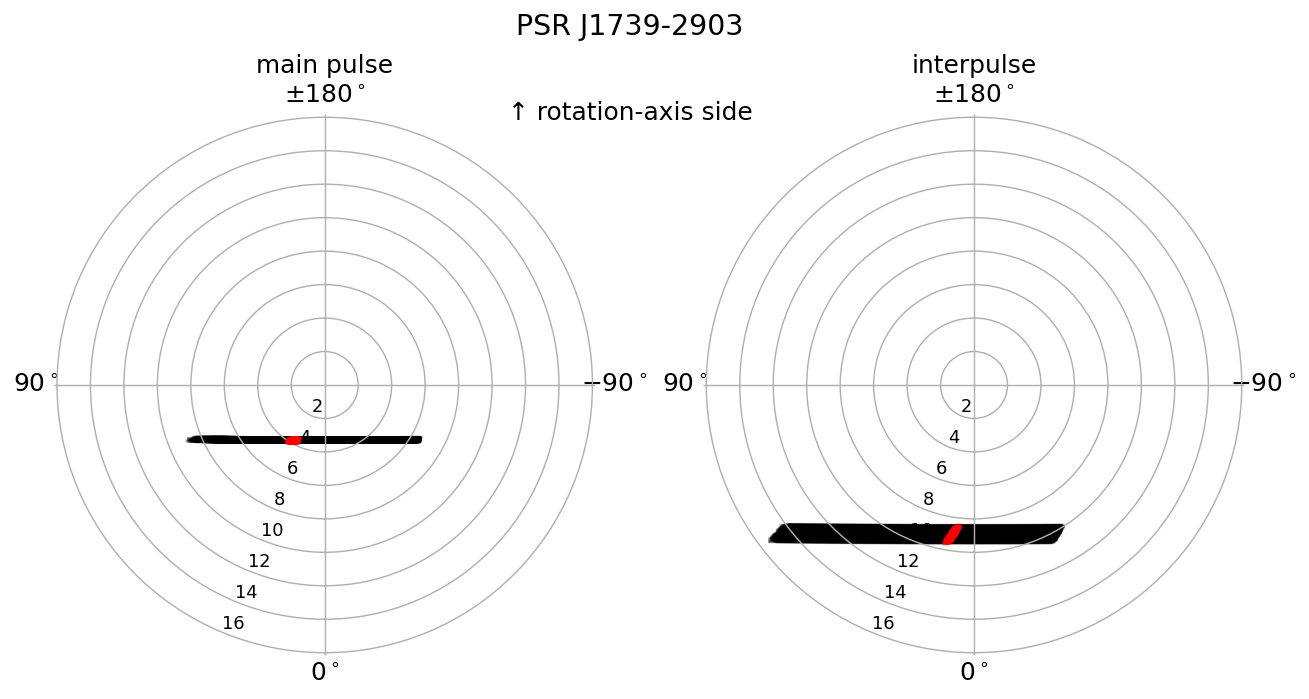}
    }
    \subfigure{
    \centering
    \includegraphics[width=0.65\textwidth]{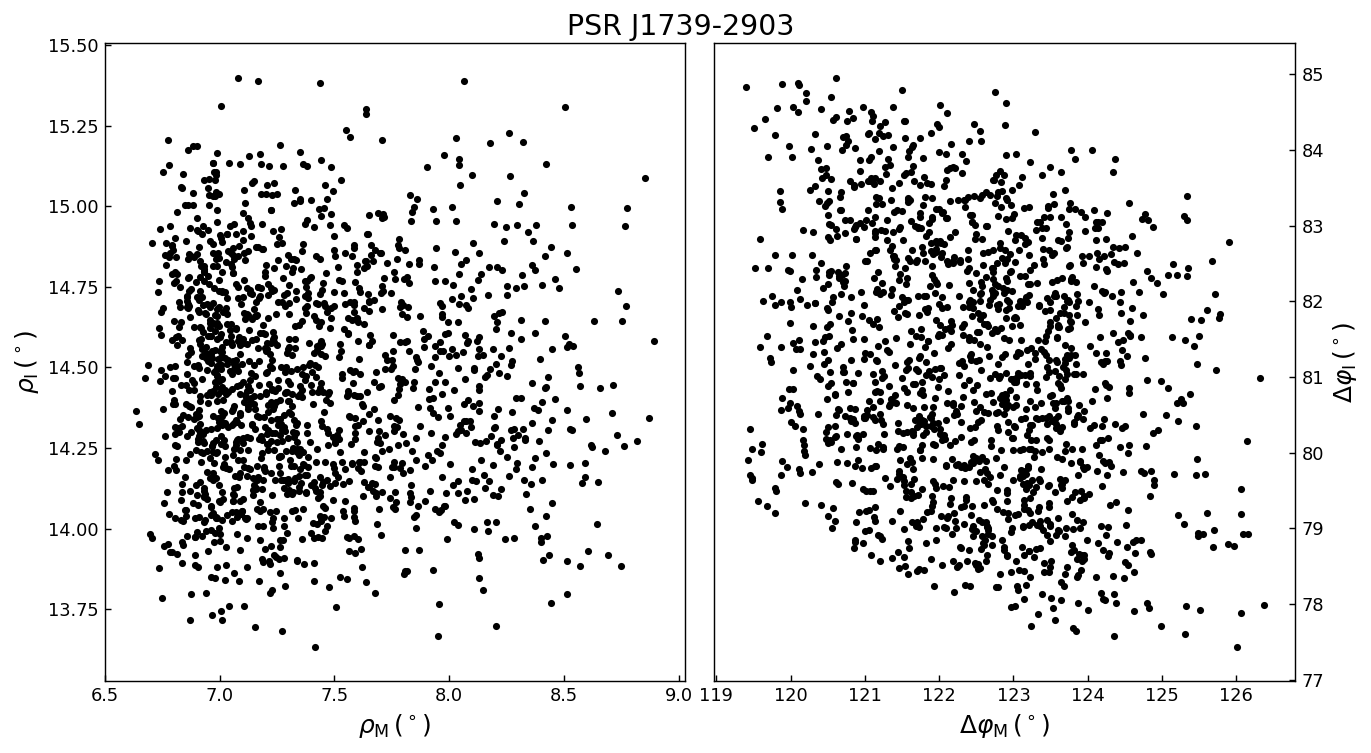}
    }
    \subfigure{
    \centering
    \includegraphics[width=0.35\textwidth]{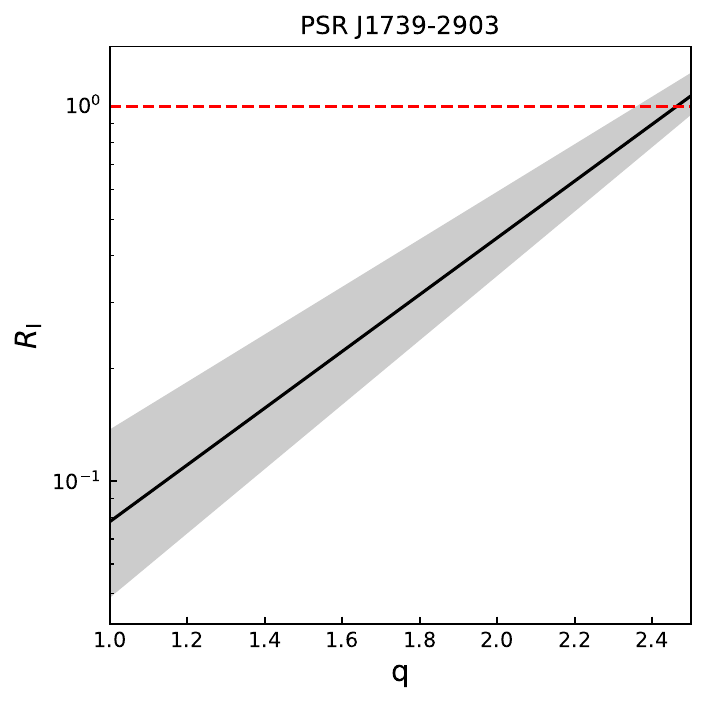}
    }
    \caption{The same as Figure \ref{fig:comp0908}, but for PSR J1739$-$2903.}
    \label{fig:comp1739}
\end{figure*}

\begin{figure*}[h]
    \centering
    \subfigure{
    \centering
    \includegraphics[width=0.65\textwidth]{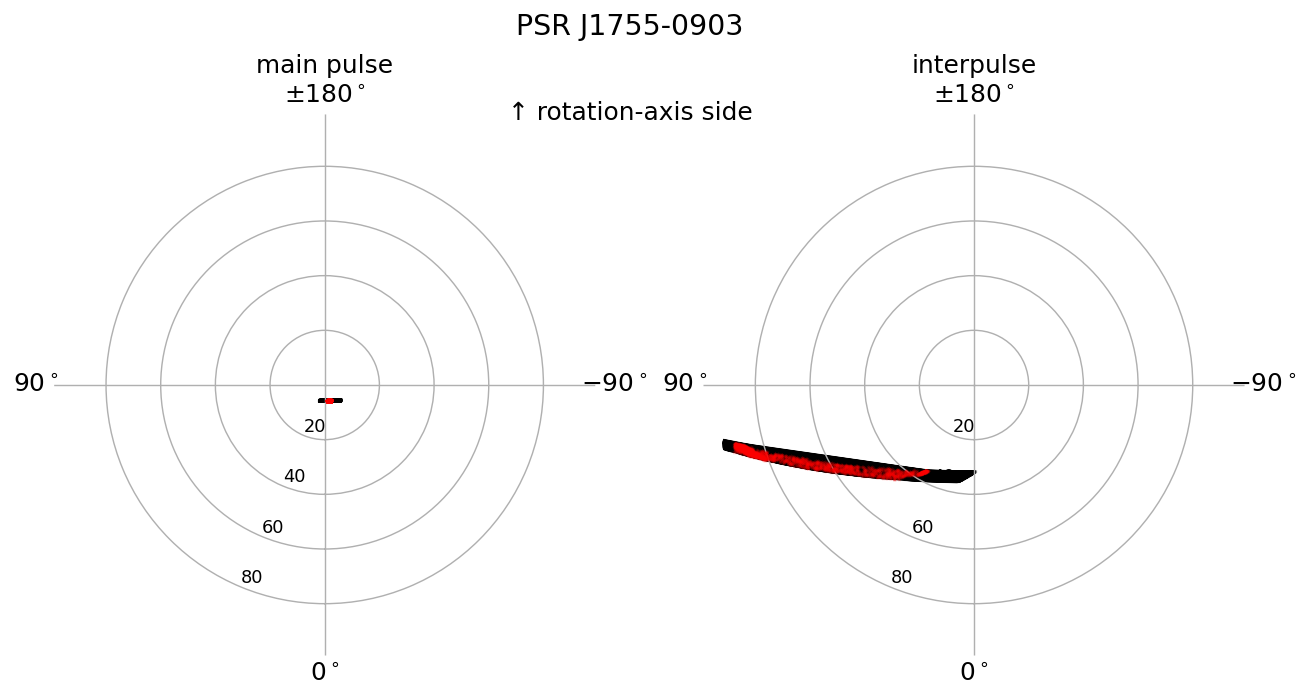}
    }
    \subfigure{
    \centering
    \includegraphics[width=0.65\textwidth]{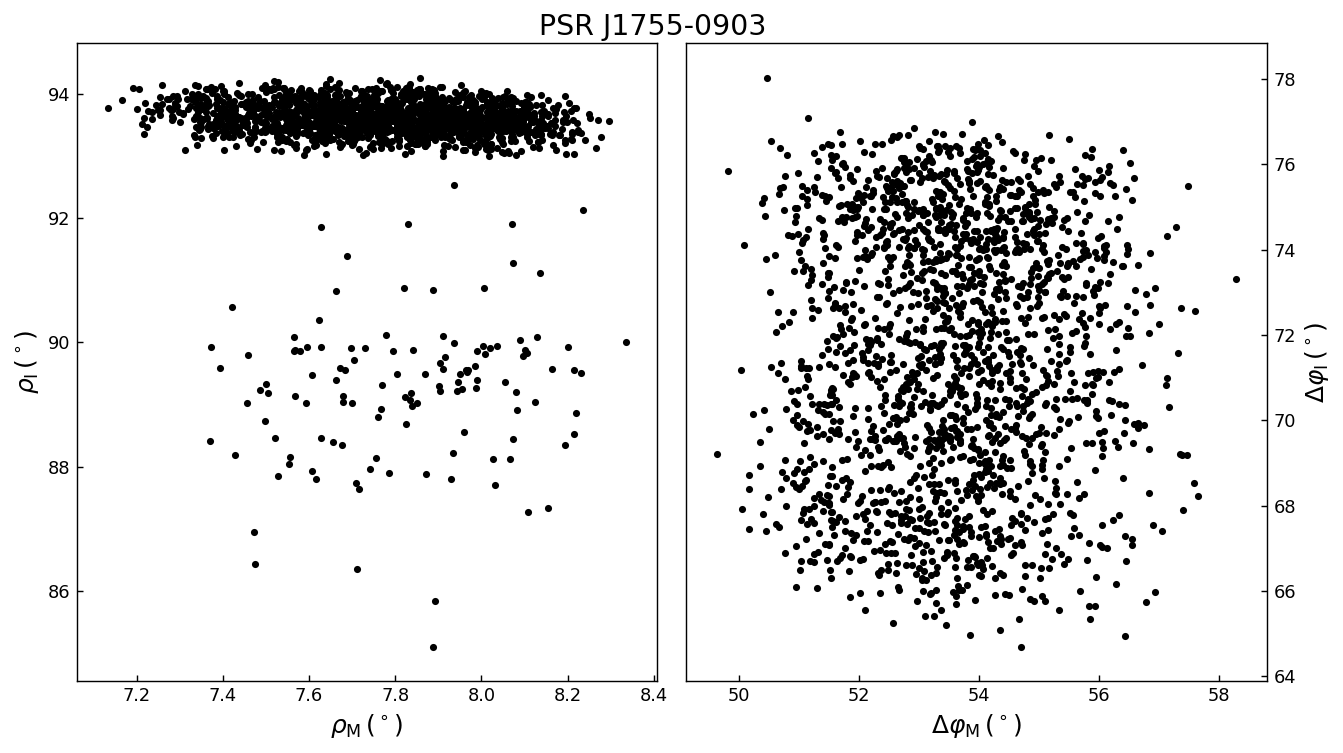}
    }
    \subfigure{
    \centering
    \includegraphics[width=0.35\textwidth]{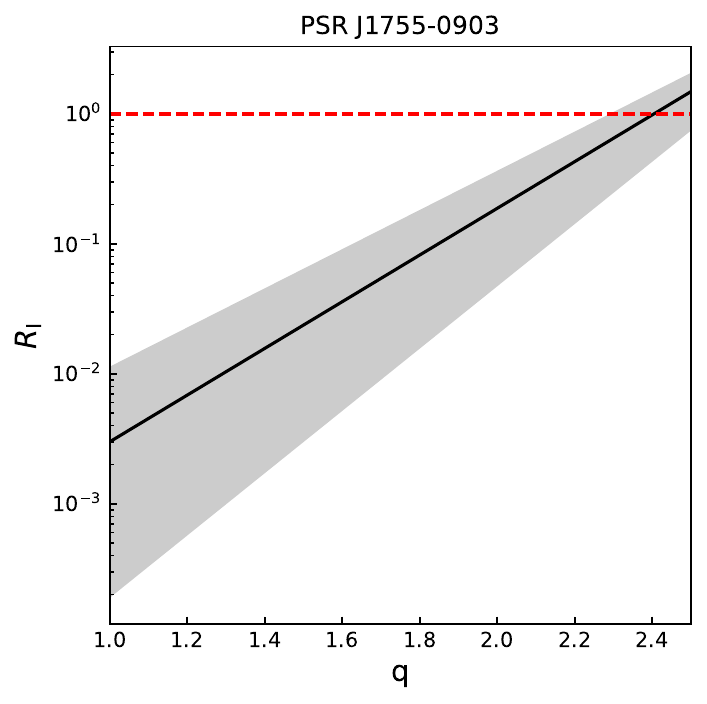}
    }
    \caption{The same as Figure \ref{fig:comp0908}, but for PSR J1755$-$0903.}
    \label{fig:comp1755}
\end{figure*}


\begin{figure*}[h]
    \centering
    \subfigure{
    \centering
    \includegraphics[width=0.65\textwidth]{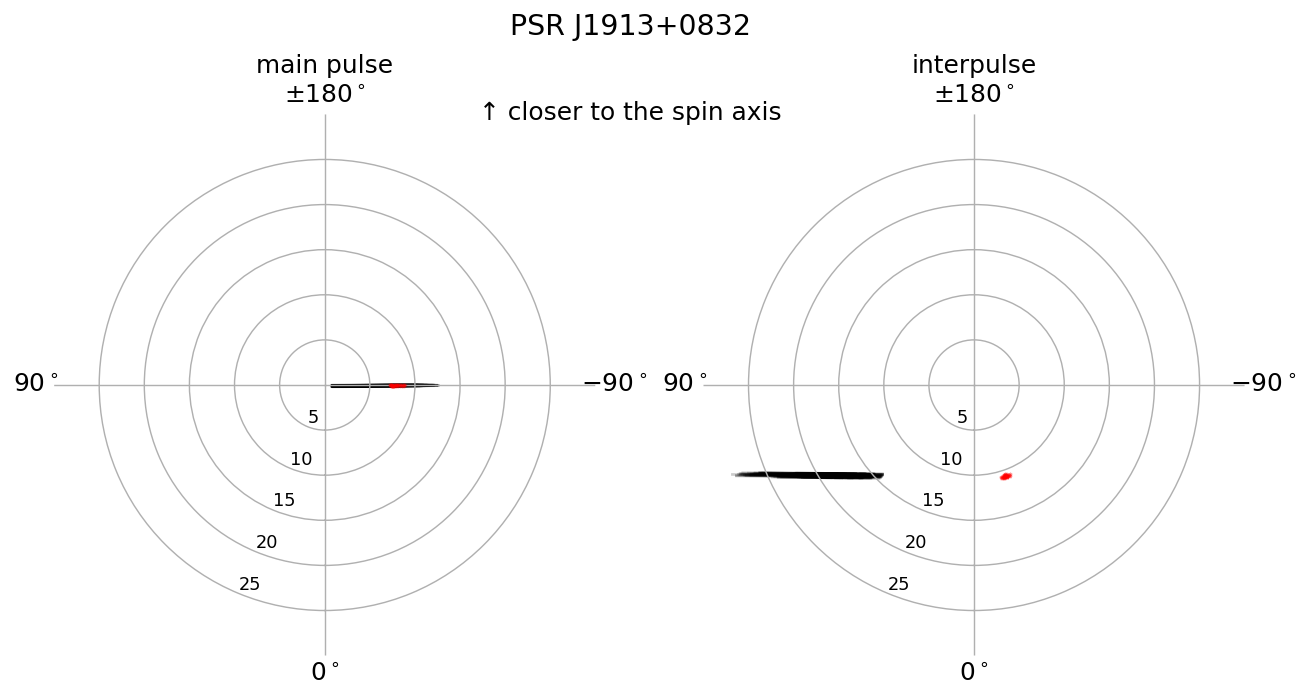}
    }
    \subfigure{
    \centering
    \includegraphics[width=0.65\textwidth]{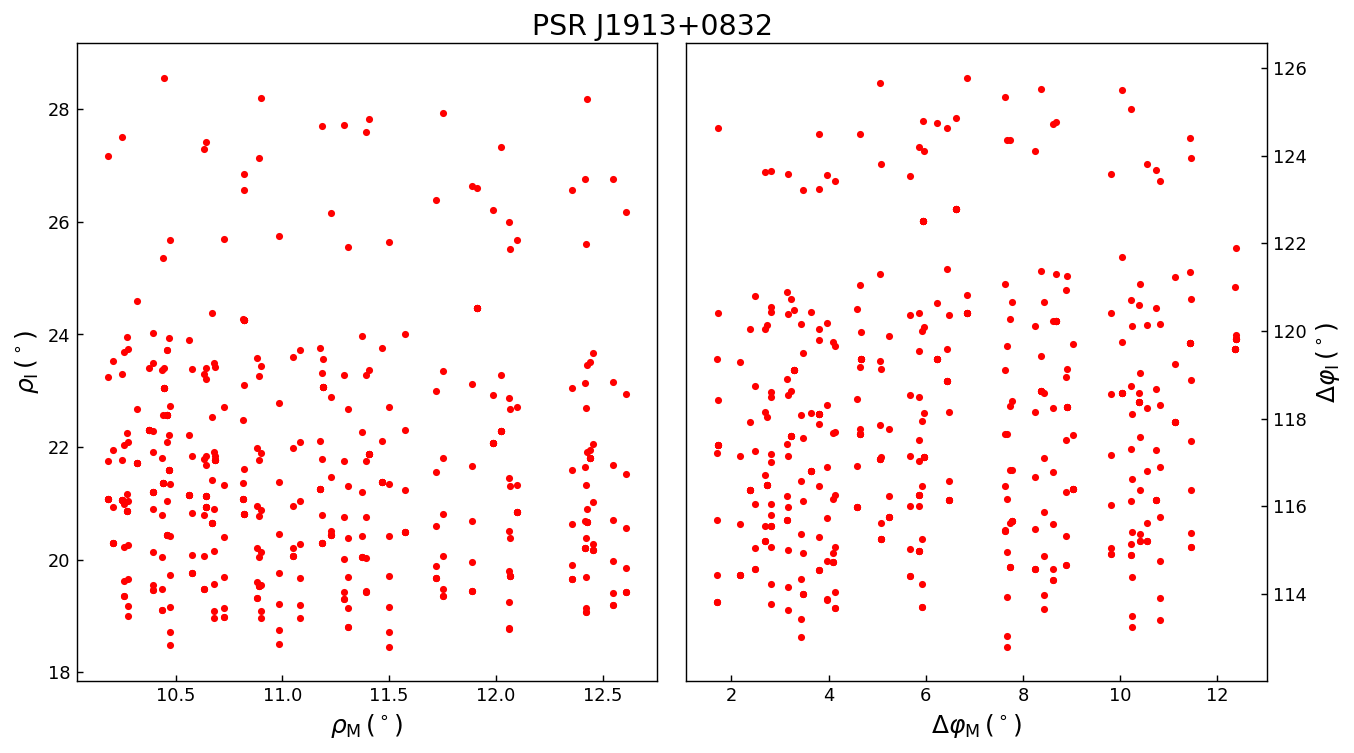}
    }
    \subfigure{
    \centering
    \includegraphics[width=0.35\textwidth]{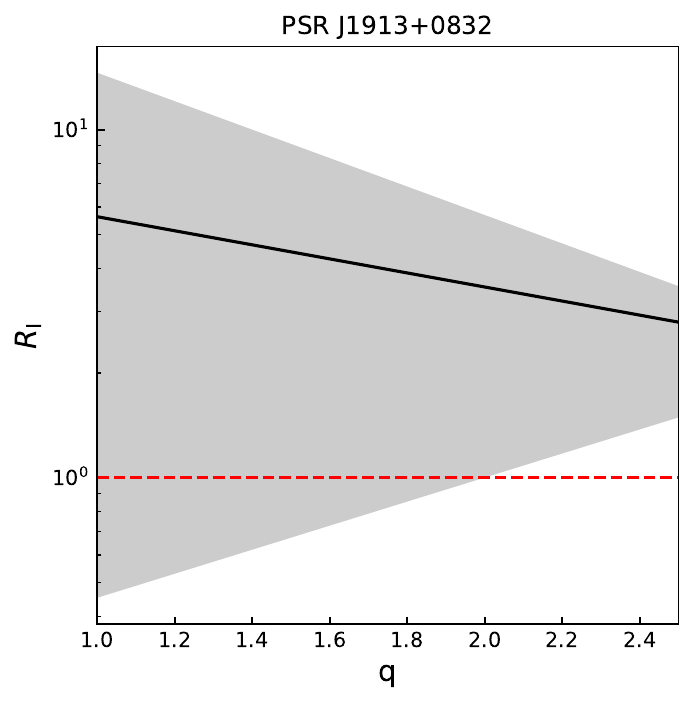}
    }
    \caption{The same as Figure \ref{fig:comp0908}, but for PSR J1909$+$0749.}
    \label{fig:comp1913}
\end{figure*}

\clearpage

\end{document}